\documentclass{aastex}
\usepackage{emulateapj5,times,mathptm}
\journalinfo{The Astronomical Journal, 2003 February, astro-ph/0211125}
\slugcomment{Received 2002 September 14; accepted 2002 November 05}

\shorttitle{SDSS Radio-Quiet Quasars: the X-ray view}
\shortauthors{VIGNALI, BRANDT, \& SCHNEIDER}

\newcommand{\ltsima}{$\; \buildrel < \over \sim \;$}
\newcommand{\simlt}{\lower.5ex\hbox{\ltsima}}
\newcommand{\gtsima}{$\; \buildrel > \over \sim \;$}
\newcommand{\simgt}{\lower.5ex\hbox{\gtsima}}

\newcommand{\lax}{{_<\atop^{\sim}}}

\newcommand{\gax}{{_>\atop^{\sim}}}
\newcommand{\cgs}{ ${\rm erg~cm}^{-2}~{\rm s}^{-1}$} 

\newcommand{\lumh}{\rm erg s$^{-1}$ Hz$^{-1}$}

\def\sun{\hbox{$\odot$}}
\def\earth{\hbox{$\oplus$}}
\def\lesssim{\mathrel{\hbox{\rlap{\hbox{\lower4pt\hbox{$\sim$}}}\hbox{$<$}}}}
\def\gtrsim{\mathrel{\hbox{\rlap{\hbox{\lower4pt\hbox{$\sim$}}}\hbox{$>$}}}}

\def\arcdeg{\hbox{$^\circ$}}
\def\arcmin{\hbox{$^\prime$}}
\def\arcsec{\hbox{$^{\prime\prime}$}}

\def\aox{$\alpha_{\rm ox}$}
\def\lo{$\log~L_{2500~\mbox{\scriptsize\rm \AA}}$}

\def\loo{{\rm $l_{uv}$}}
\def\lx{$\log~L_{\rm 2~keV}$}
\def\lxx{{\rm $l_{x}$}}
\def\bq{\tiny{$\dagger$}}

\def\chandra{{\it Chandra\/}}

\def\einstein{{\it Einstein\/}}

\def\heao1{{\it HEAO-1\/}}
\def\hst{{\it {\it HST}\/}}

\def\rosat{{\it ROSAT\/}}

\def\xmm{{XMM-{\it Newton\/}}}

\begin{document}

\title{X-RAY EMISSION FROM RADIO-QUIET QUASARS IN THE SDSS EARLY DATA RELEASE. \\
THE \aox\ DEPENDENCE UPON UV LUMINOSITY}

\author{
C. Vignali,\altaffilmark{1} 
W.~N. Brandt,\altaffilmark{1}
and D.~P. Schneider\altaffilmark{1}
}

\altaffiltext{1}{Department of Astronomy \& Astrophysics, The Pennsylvania State University, 
525 Davey Laboratory, University Park, PA 16802 \\
({\tt chris@astro.psu.edu, niel@astro.psu.edu, and dps@astro.psu.edu}).}

%\date{\today}

\begin{abstract}

We investigate the X-ray properties of the color-selected radio-quiet quasars (RQQs) 
in the Sloan Digital Sky Survey (SDSS) Early Data Release using 
\rosat, \chandra\, and \xmm\ data. 
In the 0.16--6.28 redshift range, 136 RQQs have X-ray detections 
(69 from the \rosat\ All-Sky Survey, RASS), 
while for 70 RQQs \hbox{X-ray} upper limits are obtained. 
The well-defined selection method utilized by the SDSS, coupled with 
the tight radio constraints from the FIRST/NVSS surveys, allow us to define a 
representative sample of optically selected RQQs whose broad-band spectral energy distributions 
(characterized by means of the optical-to-X-ray spectral index, \aox) 
can be studied as a function of rest-frame ultraviolet (UV) luminosity and redshift. 
A partial correlation analysis applied to the SDSS sample (including the 
upper limits, but excluding the biased subsample of RASS detections) 
shows that \aox\ is a function of rest-frame UV luminosity (i.e., \aox\ steepens at high UV luminosities); 
this correlation is significant at the 3.7$\sigma$ level. 
We do not detect a highly significant redshift dependence of \aox. 
We also find a significant (7.8$\sigma$ level) correlation between UV and X-ray luminosity. 
This correlation, parameterized by $L_{\rm X}\propto~L_{\rm UV}^{\,0.75\pm{0.06}}$, 
extends previous results to the highest redshifts. 
\end{abstract}

\keywords{galaxies: active --- galaxies: nuclei --- quasars: general --- 
X-rays: galaxies --- surveys}

\section{Introduction}

To date, many studies have focused on 
the dependence of quasar ultraviolet-to-X-ray spectral energy distributions 
[SEDs, often characterized by means of the spectral 
index \hbox{\aox=0.384\ $\log(L_{2~\rm keV}/L_{2500~\mbox{\scriptsize\AA}}$)}] 
on luminosity and cosmic time. 
Quasars are known to exhibit strong luminosity evolution in the optical band 
(e.g., Boyle et al. 1994, 2000) and perhaps luminosity-dependent density evolution in the X-ray band 
(e.g., Miyaji, Hasinger, \& Schmidt 2000; La Franca et al. 2002). 
A dependence of \aox\ on redshift would suggest 
different evolution in the optical and X-ray regimes. 

Most previous studies agree that \aox\ is only marginally dependent upon redshift, 
the primary dependence of \aox\ being on 2500~\AA\ luminosity 
(e.g., Avni \& Tananbaum 1982, 1986; Marshall et al. 1984; Kriss \& Canizares 1985; 
Tananbaum et al. 1986; Anderson \& Margon 1987; Wilkes et al. 1994; Pickering, Impey, \& Foltz 1994; 
Avni, Worrall, \& Margon 1995). 
The only significant exception has been recently reported by Bechtold et al. (2002), 
who claimed that \aox\ depends primarily upon redshift. 
These studies were generally characterized by 
heterogeneous selection criteria or covered limited ranges of redshift or luminosity. 
Most investigations included both radio-quiet quasars (RQQs) and radio-loud quasars (RLQs). 
The \hbox{X-ray} properties of these two sub-classes of the quasar population 
are now well known to be different, 
with the RLQ emission often being dominated by the jets and enhanced by boosting 
%giving rise typically to flatter SEDs in the X-ray band; 
(e.g., Wilkes \& Elvis 1987; Worrall et al. 1987; Cappi et al. 1997). 
As discussed in $\S$4.1.2.2 of Anderson (1985) and Appendix~A of Anderson \& Margon (1987), 
analyses of heterogeneous samples (such as RQQs and RLQs combined) with survival analysis techniques 
can lead to misleading results. 
Moreover, many previous samples also included Broad Absorption Line Quasars (BALQSOs). 
Recent studies (e.g., Gallagher et al. 2001, 2002; Green et al. 2001) have shown that, 
although BALQSOs are probably characterized by the same underlying X-ray continua as 
the majority of the quasar population, their X-ray emission is often depressed by 
large amounts of intrinsic absorption. 
This effect appears to apply to both low- and high-redshift BALQSOs 
(e.g., Brandt, Laor, \& Wills 2000; Brandt et al. 2001; Vignali et al. 2001a; Gallagher et al. 2002). 
Although BALQSOs do not contribute greatly to the overall optically selected quasar population 
(\hbox{$\approx$~10--15\%}; e.g., Weymann et al. 1991; 
$\S$8.1 of Brandt et al. 2000; Tolea, Krolik, \& Tsvetanov 2002), 
their inclusion can provide an ``artificial'' steepening of \aox\ as well as 
a larger, spurious scatter in any correlation study involving this spectral index 
(see Figure~4.1.2.2.a and the associated text in Anderson 1985). 

Motivated by these considerations, we have started a project to investigate 
the dependence of \aox\ upon UV luminosity and redshift 
using the quasars discovered by 
the Sloan Digital Sky Survey (SDSS; York et al. 2000) and presently published mainly 
in the Early Data Release (EDR; Stoughton et al. 2002) 
quasar catalog\footnote{Available at http://archive.stsci.edu/sdss/quasars/edrqso.cat.} 
(Schneider et al. 2002). The SDSS provides a large and well-defined sample of 
optically selected quasars with a broad range in UV luminosity and redshift, 
suitable for such studies and to break degeneracies in the 
luminosity-redshift parameter space. 
We have focused on the RQQs, since they comprise the bulk of the quasar population 
(\hbox{$\approx$~85--90\%}; e.g., Stern et al. 2000; Ivezic et al. 2002) and are not strongly 
affected by jet emission or boosting effects. 
For most of the objects, \hbox{X-ray} information has been obtained through analysis 
of \rosat\ data. For the highest redshift (i.e., $z>4$) quasars, however, 
we have also exploited the excellent capabilities of \chandra\ and \xmm. 

Throughout this paper we adopt $H_{0}$=70 km s$^{-1}$ Mpc$^{-1}$ in a $\Lambda$-cosmology 
with $\Omega_{\rm M}$=0.3 and $\Omega_{\Lambda}$=0.7 (e.g., Lineweaver 2001). 
Note that most of the \aox--\lo\ correlation papers previously cited 
%(e.g., Avni \& Tananbaum 1982, 1986; Marshall et al. 1984; Kriss \& Canizares 1985; Tananbaum et al. 1986) 
used a $H_0$=50 km s$^{-1}$ Mpc$^{-1}$ and $q_0$=0.0 cosmology, 
which provides larger luminosities by 
factors of \hbox{$\approx$~2.0--6.5} in the redshift range $\approx$~0--6.3.

\section{The sample: The SDSS Early Data Release Quasar Catalog}

The SDSS Early Data Release (EDR; Stoughton et al. 2002) 
contains 462 deg$^{2}$ of imaging data in five bands 
(designated $u$, $g$, $r$, $i$, and $z$; 
Fukugita et al. 1996; Gunn et al. 1998) and 54,008 spectra in the same area. 
The data were acquired in three regions: 
along the celestial equator in the southern Galactic sky, along the celestial equator in the 
northern Galactic sky, and in a region overlapping 
the SIRTF First Look Survey.\footnote{See http://sirtf.caltech.edu/SSC/fls.}
The SDSS EDR catalog of quasars consists of 3814 objects 
(3000 discovered by the SDSS) selected through a multicolor technique 
(similar to that described by Richards et al. 2002) 
down to magnitude limits of $i\approx$~19 and $i\approx$~20 for quasars below and above $z\approx3$, respectively 
(Stoughton et al. 2002), with an overall expected completeness higher than 90\%. 
In addition to the multicolor selection, a small percentage ($\approx$~0.5\%) 
of SDSS targets were unresolved objects brighter than $i\approx$~19 that were coincident 
with FIRST radio sources (Becker, White, \& Helfand 1995). 
The EDR quasars were selected in three stripes (each $\approx$~2.5\degr\ wide) 
over a slightly larger area \hbox{(494 deg$^{2}$)} than that covered by imaging observations, because some 
additional ``interesting'' fields were included (Stoughton et al. 2002). 
These quasars have at least 
one emission line with a full width at half maximum (FWHM) larger than \hbox{1000~km s$^{-1}$}, 
luminosities 
higher than $M_{\rm i}$=$-$23 (for $H_0$=50 km s$^{-1}$ Mpc$^{-1}$, $q_0$=0.5), 
and highly reliable redshifts ranging from 0.15 to 5.03.  
The EDR quasars used in this paper are those having the ``quasar target flag'' set to 1 in 
the EDR quasar catalog (column~22 in the quasar catalog; see $\S$4 of Schneider et al. 2002). 
In this paper the SDSS EDR sample of quasars is extended to higher redshifts by 
including all of the published SDSS quasars up to $z=6.28$ (e.g., Fan et al. 2001a) to place 
better constraints on the relevant correlations. 
The additional quasars represent only a small addition to the main EDR sample ($\approx$~2.6\%); 
since they have been selected using a multicolor selection technique, 
we do not expect any significant bias to be introduced into the analysis.

\subsection{Selection of radio-quiet quasars}

The minority of EDR quasars selected based on their detection by FIRST (see $\S$2) were immediately 
excluded from our study. 
Then from the color-selected SDSS quasars we have excluded all the RLQs, i.e., those quasars having a 
radio-loudness parameter,\footnote{The radio loudness is parameterized by 
$R$ = $f_{\rm 5~GHz}/f_{\rm 4400~\mbox{\scriptsize\AA}}$ (rest frame). 
The rest-frame 5~GHz flux density is computed from the observed-frame 1.4~GHz flux density 
assuming a radio power-law slope of $\alpha=-0.8$, with $f_{\nu}\propto~\nu^{\alpha}$.}
$R$, larger than 10 (e.g., Kellermann et al. 1989). 
Radio flux densities at 1.4~GHz have been obtained either from FIRST or from 
NVSS (Condon et al. 1998), down to typical 3$\sigma$ upper limits of 0.5 and 1.5~mJy, respectively. 
All of the quasars with \hbox{X-ray} coverage also have NVSS/FIRST coverage. 
Quasars characterized by loose radio-loudness upper limits (i.e., $R\simlt18$; only five in the sample 
of 206 SDSS quasars used in the following analysis) have been assumed to be radio quiet. 
Within the sample of SDSS quasars with \hbox{X-ray} coverage, 
the fraction of RQQs is $\approx$~90\%, which is in good agreement with 
previous estimates for optically selected quasars (e.g., Stern et al. 2000) and suggests that 
assuming the quasars with loose radio-loudness upper limits to be radio-quiet is 
reasonable.

\subsection{Optical magnitudes}

The magnitudes reported in the EDR catalog are generally accurate to 0.05 mag or better. 
To derive absolute luminosities for the SDSS quasars, the catalog 
magnitudes were first corrected for Galactic absorption 
using the {\it COBE}/DIRBE maps of Schlegel, Finkbeiner, \& Davis (1998) and the 
scaling laws for the five SDSS filters reported in Schneider et al. (2002). 
The composite quasar 
spectrum presented in Vanden Berk et al. (2001), which was constructed 
using SDSS spectra of over 2000 quasars, was used to convert the 
broad-band {\it ugriz} measurements to flux densities at rest wavelengths of 
2500~\AA\ (required for \aox\ measurements) 
and 4400~\AA\ (required to compute the radio-loudness parameter);  
this method accounts for emission lines as they pass through the SDSS filters. 
The offsets for both the rest-wavelength 
2500~\AA\ and 4400~\AA\ flux densities as a function of redshift 
were calculated for each of the SDSS filters, using interpolation whenever possible. 
For example, if the observed wavelength of rest-frame 2500~\AA\ occurred between the $g$ and 
$r$ filters, the value of the 2500~\AA\ flux density was calculated using the relations for both 
the $g$ and $r$ filters, and the final measurement was 
the weighted average of the 2500~\AA\ flux densities, 
weighted by the relative distances of $2500(1+z)$~\AA\ from the central wavelengths 
of the $g$ and $r$ filters. If rest-wavelength 1280~\AA\ occurred redward 
of the central wavelength of a filter, that filter was not used 
in the calculation, due to the large dispersion in the relations introduced by 
the presence of the Lyman~$\alpha$ emission line and the Lyman~$\alpha$ forest. 
If the observed-frame wavelength of the desired region of the spectrum was larger 
than the effective wavelength of the reddest filter, 
the relation for the reddest filter was used.

\section{The X-ray data: Data reduction and analysis}

\subsection{The $z<4$ sample: \rosat\ observations}

The EDR quasar catalog was cross correlated with archival \rosat\ data 
[pointed PSPC, pointed HRI, and the \rosat\ All-Sky Survey (RASS)]. 
We used only the central regions (20\arcmin\ radius) of the \rosat\ PSPC and HRI detectors, 
where the sensitivity is highest and 
the PSPC window support structure does not affect source detection. 
This maximizes the probability of detecting faint sources. 
In cases of multiple observations of the same SDSS quasar, 
we chose the one with the best combination of off-axis angle and exposure time. 
Approximately 3.2\% of the EDR area is covered by \rosat\ pointed observations [considering only 
the inner 20\arcmin\ of the \rosat\ field-of-view; 37 PSPC fields (2 largely overlapping) and 11 HRI fields; 
see Fig.~1]. 
Source detection was performed in the \hbox{0.5--2~keV} band for the PSPC and 
in the full band for the HRI, since no spectral information is available for the HRI. 
A conservative matching radius of 40\arcsec\ was used to take into account the 
broadening of the Point Spread Function (PSF) at large off-axis 
angles.\footnote{See ftp://ftp.xray.mpe.mpg.de/rosat/catalogues/1rxp/wga\_rosatsrc.html.} 

The pointed \rosat\ PSPC and HRI data have been analyzed with the MIDAS/EXSAS package 
(Zimmermann et al. 1998). 
The sources have been detected by running the local 
detection algorithm {\sc LDETECT}, the bicubic spline fit 
to the background map, and the map detection algorithm {\sc MDETECT}. 
%
%%%%%%%%%%%%%%%%%%%%%%%%%%%%%%%%%%%%%%%%%%%%%%%%%%%%%%%%%%%%%%%%%%%%%%%%%%%
%%%	FIGURE 1: Redshift and Optical Luminosity (2500 A rest frame) distributions for SDSS RQQs. 
%%%%%%%%%%%%%%%%%%
\end{multicols}
\begin{figure*}
\figurenum{1}
\vskip -0.6cm
\centerline{\includegraphics[angle=0,width=\textwidth]{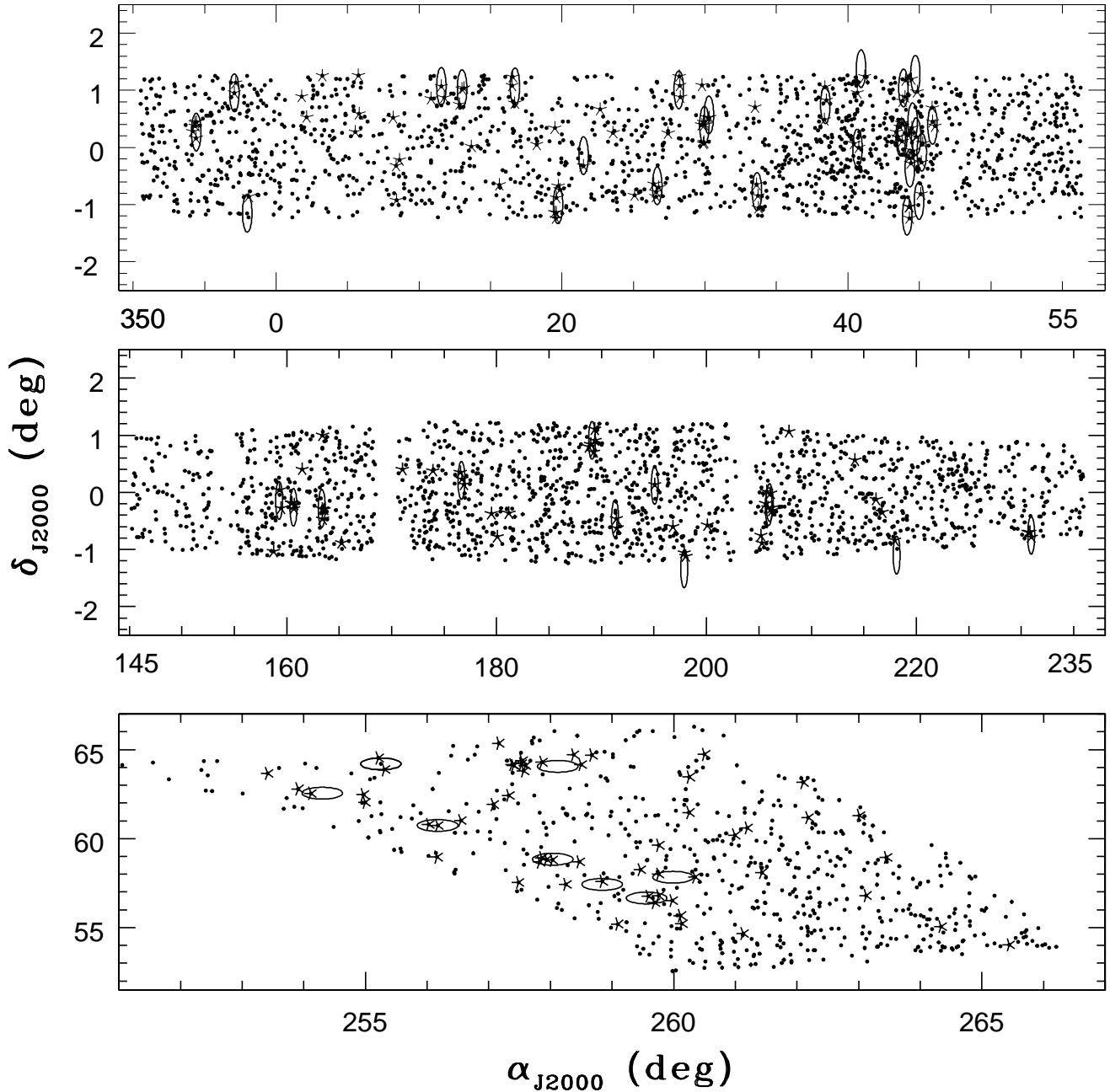}}
\figcaption{\footnotesize 
The \rosat\ coverage of the SDSS EDR quasars (small points) 
with pointed PSPC and HRI observations (circles of radius 20\arcmin\ in sky coordinates). 
The stars indicate the SDSS EDR RQQs used in this work. The small points in some circles are RLQs, 
which were not used in this work. 
\label{fig1}}
\vglue-0.2cm
\end{figure*}
\begin{multicols}{2}
%%%%%%%%%%%%%%%%%%
%%%	END of FIG.1
%%%%%%%%%%%%%%%%%%%%%%%%%%%%%%%%%%%%%%%%%%%%%%%%%%%%%%%%%%%%%%%%%%%%%%%%%%%
%
%%%%%%%%%%%%%%%%%%%%%%%%%%%%%%%%%%%%%%%%%%%%%%%%%%%%%%%%%%%%%%%%%%%%%%%%%%%
%%%	TABLE 1: ROSAT, Chandra, and XMM-Newton Results
%%%%%%%%%%%%%%%%%%
\begin{table*}[!ht]
\footnotesize
\caption{Numbers of SDSS RQQs in \rosat, \chandra, and \xmm\ observations}
\begin{center}
\begin{tabular}{lcccccccccc}
\hline
\hline
Redshift & \multicolumn{5}{c}{\rosat} & \multicolumn{1}{c}{} & \multicolumn{2}{c}{\chandra} & 
\multicolumn{1}{c}{} & \multicolumn{1}{c}{\xmm} \\
Range & \multicolumn{10}{c}{} \\
\cline{2-6} \cline{8-9} \cline{11-11} \\
  & \multicolumn{2}{c}{PSPC} & \multicolumn{2}{c}{HRI} & \multicolumn{1}{c}{RASS} 
& \multicolumn{1}{c}{} & \multicolumn{2}{c}{ACIS} & \multicolumn{1}{c}{} & \multicolumn{1}{c}{EPIC} \\
  & \multicolumn{1}{c}{D} & \multicolumn{1}{c}{UL} & 
\multicolumn{1}{c}{D} & \multicolumn{1}{c}{UL} & \multicolumn{1}{c}{D} & \multicolumn{1}{c}{} & 
\multicolumn{1}{c}{D} & \multicolumn{1}{c}{UL} & \multicolumn{1}{c}{} & \multicolumn{1}{c}{D} \\
\hline
$z<2$   & 43 & 32 & {\phn}5 & 19 & 68 & & \nodata & \nodata & & \nodata \\
$2<z<4$ & 11 & 10 & {\phn}2 & {\phn}6 & {\phn}1 & & \nodata & \nodata & & \nodata \\
$z>4$   & \nodata & \nodata & \nodata & \nodata & \nodata & & {\phn}5 & {\phn}3 & & {\phn}1 \\
\hline
\end{tabular}
\vskip 2pt
\parbox{0.65\textwidth}
{\small\baselineskip 9pt
\footnotesize
\indent
{\sc Note. ---}
``D'' indicates the number of X-ray detections, while ``UL'' indicates the number of upper limits.
}
\end{center}
\vglue-0.75cm
\label{tab1}
\end{table*}
\normalsize
%%%%%%%%%%%%%%%%%%%%%
%%%	End of Table 1
%%%%%%%%%%%%%%%%%%%%%%%%%%%%%%%%%%%%%%%%%%%%%%%%%%%%%%%%%%%%%%%%%%%%%%%%%%%
%
The detection threshold of these algorithms was set at a likelihood of 
$L=-\ln(P_{\rm e})=6$, corresponding to a probability $P_{\rm e}$ of the order of 2.5$\times$10$^{-3}$ 
that the observed number of photons in the source cell is produced entirely by a 
background fluctuation (corresponding to the $\approx$ 3$\sigma$ detection level; 
Cruddace, Hasinger, \& Schmitt 1988). 
In the present sample, the sources detected by the \rosat\ PSPC (HRI) have typical offsets of less 
than \hbox{15--20\arcsec} (10\arcsec) from their optical positions, 
which are accurate to within 0.2\arcsec\ (Schneider et al. 2002). 
This range of offsets is consistent with the values found 
recently by Vignali et al. (2001b) for faint hard X-ray selected sources in PSPC/HRI fields. 
All of the sources were inspected by eye to verify that they are not spurious 
fluctuations due to enhanced local background. 
Our previous experience (e.g., Vignali et al. 2001b) suggests that the combination of detection 
algorithms described above is quite effective at minimizing the number of spurious detections. We expect 
only $\approx$~0.3 false sources in our sample due to background 
fluctuations adopting the detection threshold reported above. 
Assuming the integral 0.5--2~keV source counts of Hasinger et al. (1998), 
we conservatively expect $\simlt2$ spurious associations within the present sample of 
\rosat-detected sources. 
%
% <3 including also the RASS sources which are detected at higher soft X-ray fluxes (so <<3). 
%

The background-subtracted, vignetting-corrected source counts were 
determined using the maximum likelihood method ({\sc MAXLIK}). 
For the sources not detected in the \rosat\ pointed observations, 3$\sigma$ X-ray upper limits 
were computed using the {\sc SOSTA} task in the {\sc XIMAGE} package 
(Giommi et al. 1992) and an average value for the background close to the quasar position. 

The same matching radius was utilized for the RASS data (see, e.g., Voges et al. 1999) 
in the \hbox{0.1--2.4~keV} band with a slightly higher detection threshold [$L=-\ln(P_{\rm e})=7$, 
the minimum available using the \rosat\ 
Source Browser\footnote{Available at http://www.xray.mpe.mpg.de:80/cgi-bin/rosat/src-browser.}]. 
The average exposure time for the quasars detected in the RASS is $\approx$~1.2~ks. 
% Range RASS EXPO: 220-3750 s

\subsubsection{\rosat\ results}

The results of the cross correlation of the SDSS EDR quasars with 
archival \rosat\ data are presented 
in Table~1. 
%
%%%%%%%%%%%%%%%%%%%%%%%%%%%%%%%%%%%%%%%%%%%%%%%%%%%%%%%%%%%%%%%%%%%%%%%%%%%
%%%	FIGURE 2: Redshift and Optical Luminosity (2500 A rest frame) distributions for SDSS RQQs. 
%%%%%%%%%%%%%%%%%%
%\begin{figure}
\figurenum{2}
\centerline{\includegraphics[angle=0,width=8.5cm]{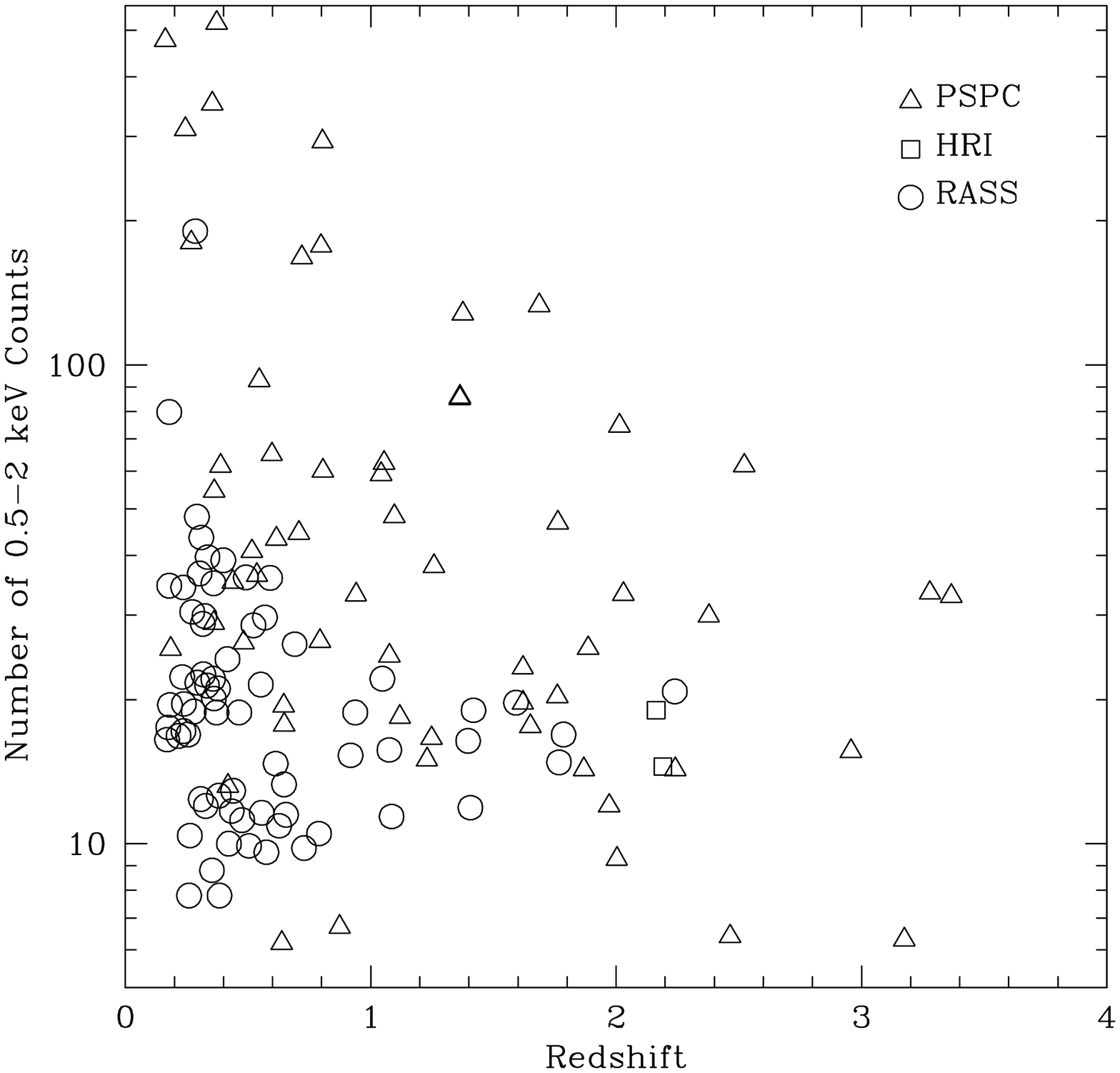}}
\figcaption{\footnotesize 
Number of counts (corrected for vignetting) in the 0.5--2~keV band for \rosat--detected SDSS 
RQQs as a function of redshift. 
Triangles, squares, and circles represent pointed PSPC, pointed HRI, and RASS detections, respectively. 
\label{fig2}}
\centerline{}
\centerline{}
%\end{figure}
%%%%%%%%%%%%%%%%%%
%%%	END of FIG.2
%%%%%%%%%%%%%%%%%%%%%%%%%%%%%%%%%%%%%%%%%%%%%%%%%%%%%%%%%%%%%%%%%%%%%%%%%%%
%
None of the SDSS EDR quasars at $z>4$ lies in the inner 20\arcmin\ of a \rosat\ field. 
We found 54 X-ray detected RQQs in pointed PSPC fields, seven in pointed HRI fields, and 69 in the RASS. 
The numbers of upper limits were 42 and 25 for the PSPC and HRI, respectively. 
Only a small fraction ($\approx$~4.7\%) of the \rosat\ pointed observations were targeting SDSS quasars 
(five detections and one upper limit); 
the relevant quasars are listed in Table~2 with a value of 0.0\arcmin\ for the off-axis angle. 
The exclusion of the \rosat\ targets from the following analyses does not change materially any of the results 
presented below. 
A plot of the number of counts from \rosat-detected quasars as a function of redshift is 
shown in Fig.~2. 
Most of the X-ray detections are too faint to derive tight spectral constraints 
by means of hardness ratio analysis. 
Ten quasars have less than 10 counts in the \hbox{0.5--2~keV} band 
(seven quasars at $z<2$, two observed by the PSPC and five by the RASS; three at $2<z<4$, all observed by the PSPC). 
All of these faint objects satisfy the detection threshold because of 
the high sensitivity generally reached in the inner part of \rosat\ field-of-view 
and the low background in some cases. These sources appear real after visual inspection, and in two out of five cases 
we were able to confirm the source's existence by analyzing an additional \rosat\ PSPC observation (in the other cases 
no additional pointed observations were available). 

Among the quasars with pointed \rosat\ coverage, we found seven BALQSOs; 
one was detected by \rosat\ and six were not. 
Unfortunately, no information about the presence of broad UV absorption features 
is available for $z\lax1.5$ quasars due to the wavelength range of the SDSS spectrograph 
(3800--9200~\AA; York et al. 2000). 
The relatively small fraction of BALQSOs within optically selected samples (\hbox{$\approx$~10--15\%}; e.g., 
Weymann et al. 1991; Brandt et al. 2000; Tolea et al. 2002) 
suggests that BALQSOs probably do not significantly contribute to the 
estimation of the average properties of the present sample at $z\lax1.5$. 
We have searched in the NASA Extragalactic Database (NED) 
for published UV spectra of the SDSS quasars at $z\lax1.5$. 
A small fraction (below 1\%) of the low-redshift SDSS quasars have UV coverage, and none of these 
shows broad absorption features. 
In $\S$4.3 and $\S$4.4 we will statistically estimate the impact of BALQSOs 
on \aox\ determinations at $z\lax1.5$.

\subsection{The $z>4$ sample: \chandra\ and \xmm\ observations}

All but one of the $z>4$ SDSS quasars in the \hbox{X-ray} sample have been observed 
by \chandra, the only exception being the $z=5.74$ quasar (Fan et al. 2000b) detected by \xmm\ 
(Brandt et al. 2001). This quasar has recently been revealed to be a 
BALQSO (Maiolino et al. 2001; Goodrich et al. 2001).  
The \chandra\ results reported in this paper 
have been taken from Vignali et al. (2001a; five SDSS RQQs, two of which are X-ray undetected BALQSOs) and 
Brandt et al. (2002; two SDSS RQQs). 
\chandra\ data for one additional SDSS RQQ have been retrieved from the archive 
(also published by Bechtold et al. 2002). 
All of the \chandra\ sources were observed with the back-illuminated ACIS-S/S3 CCD 
(Garmire et al. 2002). 
Source detection was carried out with {\sc wavdetect} (Freeman et al. 2002). 
For each image, we calculated wavelet transforms (using a Mexican hat kernel) 
with wavelet scale sizes of 1, 1.4, and 2 pixels 
using a false-positive probability threshold of 10$^{-6}$ 
(see Vignali et al. 2001a for further information about the \chandra\ reduction and detection processes).

To increase the number of quasars at $z>4$ populating the high-luminosity tail of the 
luminosity function, 
and therefore to place stronger constraints on the parameters involved in this study, 
we will also include in the analysis 13 luminous RQQs taken from the 
Palomar Digital Sky Survey (DPOSS; Djorgovski et al. 1998; hereafter referred to as 
PSS) which have been observed by \chandra\ (Vignali et al. 2001a, 2003). 
Twelve of these objects are detected in the soft (0.5--2~keV) \hbox{X-ray} band. 
Given their high luminosities, these objects are sufficiently bright that 
any quasars of this class that fall in the 
SDSS survey area will be included in the SDSS spectroscopic survey 
(indeed, many PSS quasars have already been recovered by the SDSS; 
see Schneider et al. 2002 and references therein). 
Therefore, we do not expect to introduce biases by including PSS quasars in our 
otherwise pure SDSS sample. 
The critical results below will be given both with and without inclusion of the PSS quasars.

\subsubsection{\chandra\ and \xmm\ results}

A summary of the \chandra\ and \xmm\ results for $z>4$ SDSS RQQs is presented in Table~1. 
We found six X-ray detected SDSS RQQs and three upper limits, two of which 
are associated with BALQSOs (see Vignali et al. 2001a for a detailed discussion). 
The last X-ray undetected quasar is peculiar; its 
optical spectrum is characterized by a lack of emission lines 
(Fan et al. 1999, 2000a).

\subsection{Count rate-to-flux conversion}

All of the X-ray fluxes reported in this paper have been converted into the observed 0.5--2~keV band 
and corrected for Galactic absorption. 
The count rate-to-flux conversion was computed using the 
Portable, Interactive, Multi-Mission Simulator (PIMMS Version 3.2c; Mukai 2001) software 
for a power-law photon index\footnote{$\Gamma$=$-$$\alpha$$+$1; 
$N(E)\propto~E^{-\Gamma}$, where $N(E)$ has units of photons cm$^{-2}$ keV$^{-1}$ s$^{-1}$.} 
of $\Gamma=2$ and Galactic absorption (Dickey \& Lockman 1990). 
Samples of $z\approx$~0--2 RQQs are well fitted in the \hbox{2--10~keV} band 
by power-law continua with \hbox{$\Gamma$=1.7--2.3} (e.g., George et al. 2000; Mineo et al. 2000; Reeves \& Turner 2000). 
Although there have been claims of a possible flattening of \hbox{X-ray} spectral continua 
at $z\approx$~2 (e.g., Vignali et al. 1999; Blair et al. 2000), the assumption of a flatter slope  
($\Gamma=1.5$) has only a few percent effect ($\approx$~1--2\%) on the derived fluxes. 
In the same manner, assuming a steeper power-law slope ($\Gamma=2.5$) 
has a few percent effect on the derived fluxes. 
It must also be kept in mind that, 
given the large redshift range of the SDSS quasars used here, 
we are not sampling the same rest-frame \hbox{X-ray} bandpass for the entire SDSS sample; 
therefore the assumption of a single power-law model for the 
count rate-to-flux conversion may be too simplistic. 
However, although the direct spectral information available for $z>2$ high-luminosity RQQs is still 
limited, there are indications that quasars at the highest redshifts are also well parameterized by 
$\Gamma\approx$~2 power-law \hbox{X-ray} continua (e.g., Dewangan et al. 2002; Page et al. 2002; Vignali et al. 2003). 
%
%%%%%%%%%%%%%%%%%%%%%%%%%%%%%%%%%%%%%%%%%%%%%%%%%%%%%%%%%%%%%%%%%%%%%%%%%%%
%%%	FIGURE 3: 0.5-2 keV [observed frame] flux distribution for the SDSS RQQs
%%%%%%%%%%%%%%%%%%
%\begin{figure}
\figurenum{3}
\centerline{\includegraphics[angle=0,width=8.5cm]{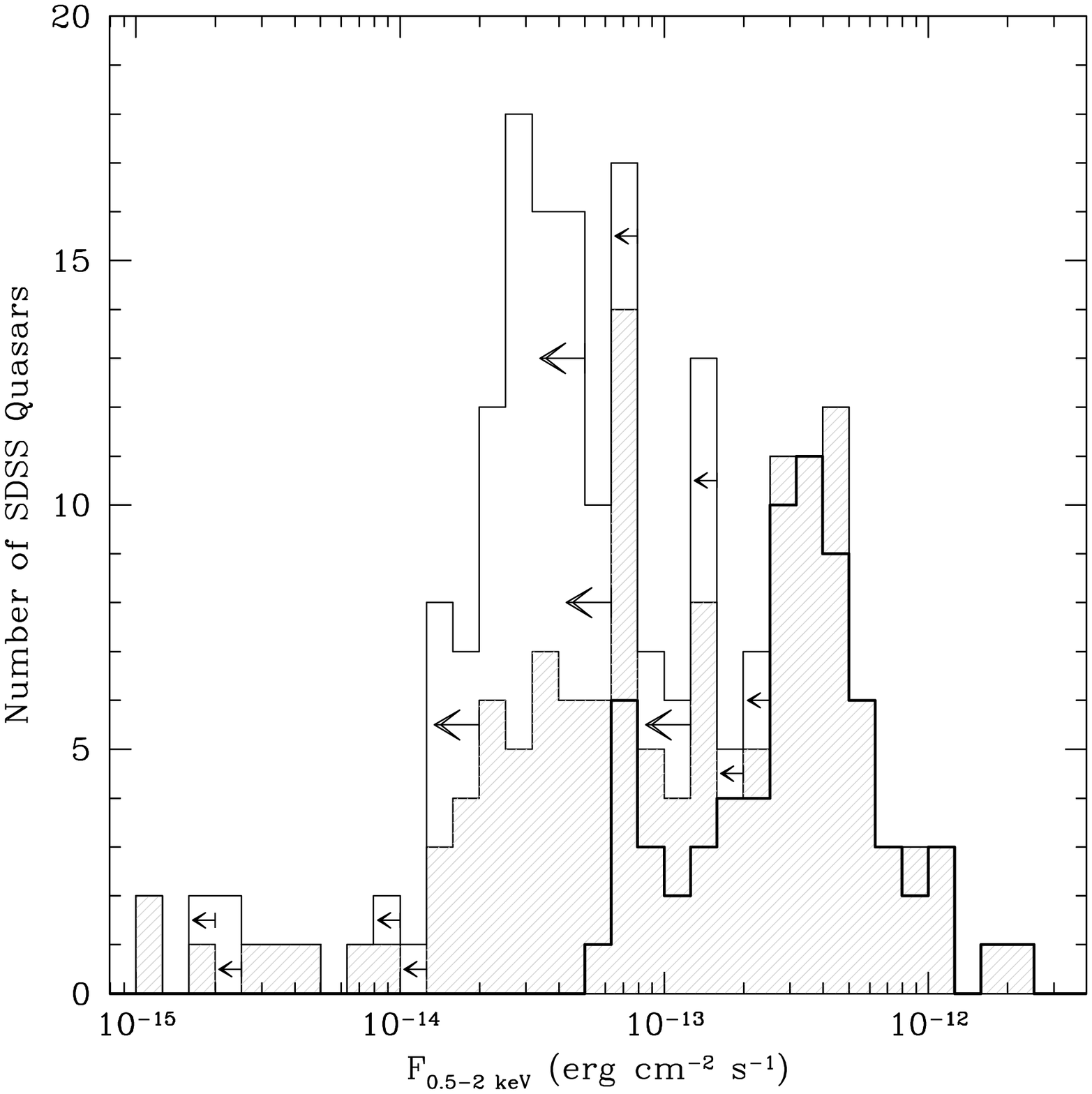}}
\figcaption{\footnotesize 
Galactic absorption-corrected 0.5--2~keV flux distribution 
for the 206 SDSS RQQs with X-ray information. 
The shaded area indicates the quasars with X-ray detections, 
the thin continuous line those from the entire sample, and 
the thick line the ones detected by the RASS. 
Unshaded areas with arrows indicate upper limits. 
\label{fig3}}
\centerline{}
\centerline{}
%\end{figure}
%%%%%%%%%%%%%%%%%%
%%%	END of FIG.3
%%%%%%%%%%%%%%%%%%%%%%%%%%%%%%%%%%%%%%%%%%%%%%%%%%%%%%%%%%%%%%%%%%%%%%%%%%%

The observed-frame \hbox{0.5--2~keV} flux distribution for our sample is plotted in Fig.~3. 
Most of the X-ray brightest objects have been detected by the RASS (thick continuous line).

\subsection{SDSS RQQs observed in the X-ray band} 

The combined SDSS, \rosat, \chandra, and \xmm\ data allow a systematic study of the 
ultraviolet-to-X-ray properties of SDSS RQQs in the redshift range \hbox{$z$=0.16--6.28}. 
A total of 206 RQQs have X-ray information: 136 have been detected, while for 70 RQQs 3$\sigma$ upper limits 
have been derived.
Systematic cross-calibration errors might be present when data from different \hbox{X-ray} instruments 
are analyzed. However, to date we have no indication of any \rosat\ versus \chandra\ 
cross-calibration problems 
(T.~Gokas 2002, private communication).\footnote{See also http://cxc.harvard.edu/cal/Links/Acis/acis/Presentations/cross\_cal/wa2-snowden.}
The principal optical, \hbox{X-ray}, and radio information for the SDSS RQQs 
with X-ray coverage is summarized in Table~2.\footnote{This table is also available 
at http://www.astro.psu.edu/users/niel/papers/papers.html.}  
In Fig.~4 we compare the redshift and rest-frame 2500~\AA\ luminosity density distributions of 
the X-ray detected SDSS quasars (shaded area) with the overall 
sample of SDSS quasars with \hbox{X-ray} information (thin continuous line). 
The thick line shows the SDSS RQQs detected by the RASS, most of which are at $z<1.5$. 
The sample has a fairly high fraction of X-ray detections 
($\approx$~49\% if only pointed observations are taken into account). 
Among the 206 SDSS RQQs, 11 objects at low redshift are optically resolved 
(Schneider et al. 2002). 
They do not populate a particular region in the 
optical--X-ray luminosity space, suggesting that the emission from their host galaxies 
is likely to be negligible. 
%
%%%%%%%%%%%%%%%%%%%%%%%%%%%%%%%%%%%%%%%%%%%%%%%%%%%%%%%%%%%%%%%%%%%%%%%%%%%
%%%	FIGURE 4: Redshift and Optical Luminosity (2500 A rest frame) distributions for SDSS RQQs. 
%%%%%%%%%%%%%%%%%%
\begin{figure*}
\figurenum{4}
\resizebox{8.5cm}{!}{\includegraphics{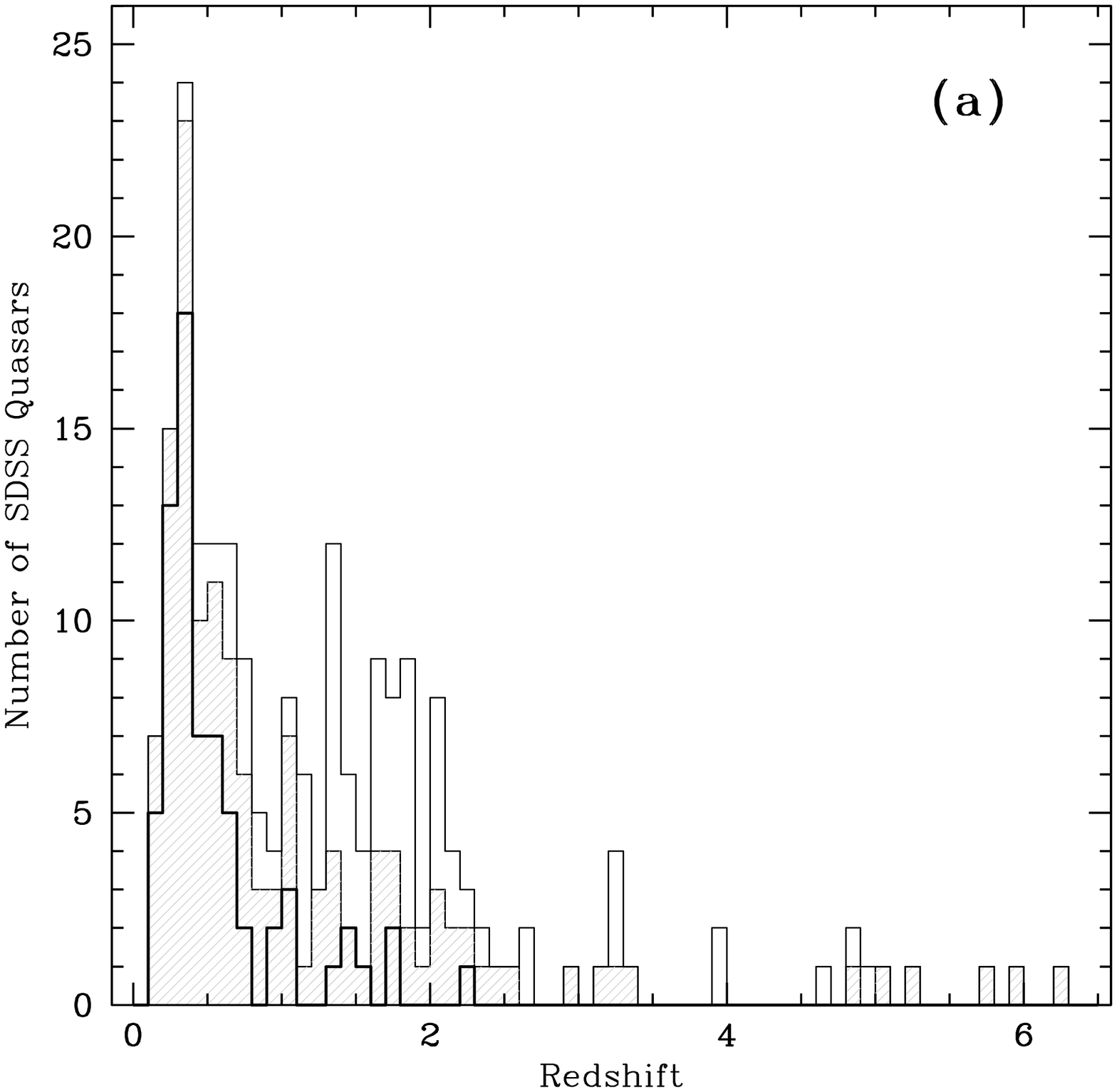}}
\hspace{1cm}
\resizebox{8.5cm}{!}{\includegraphics{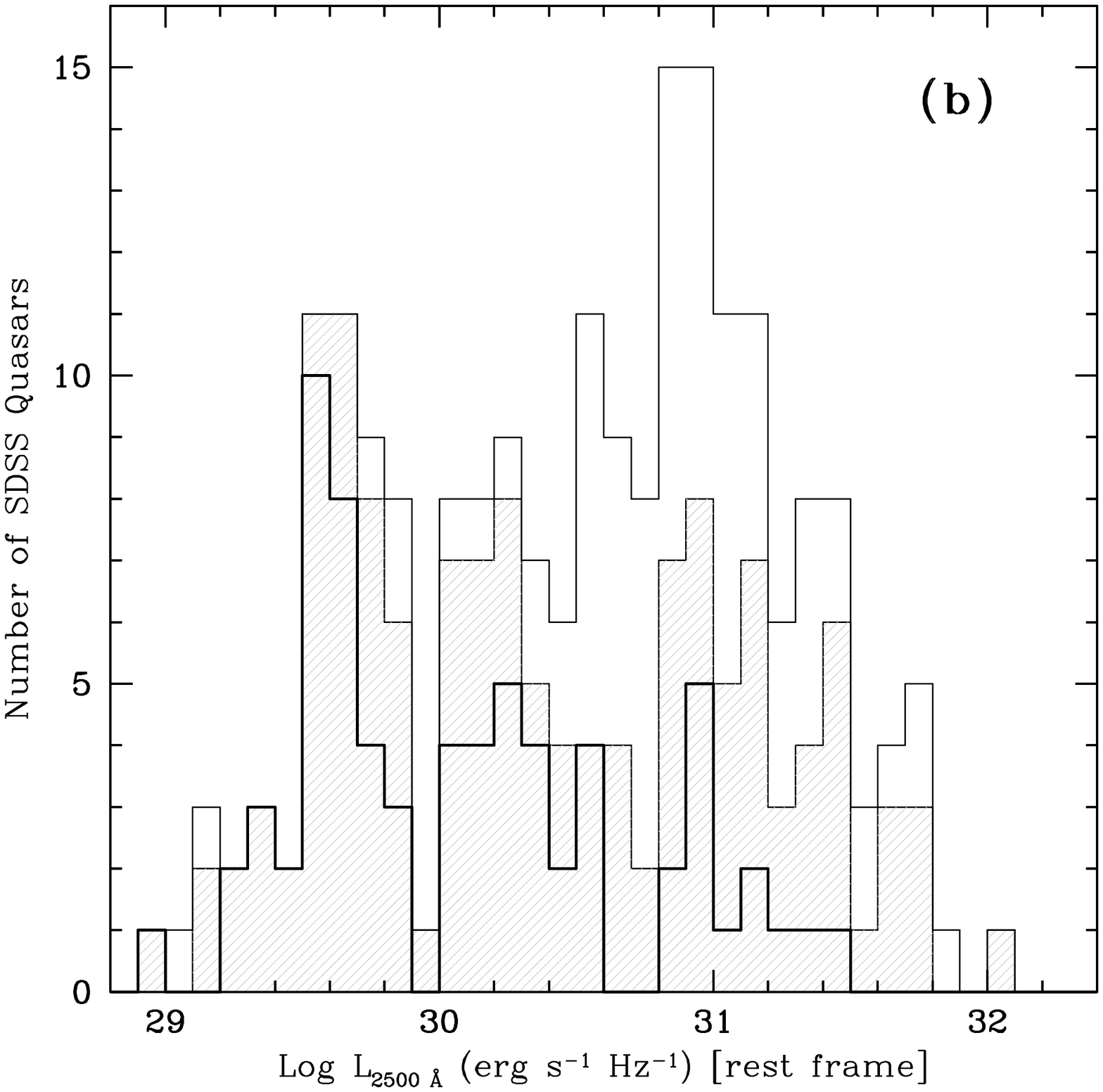}}
\figcaption{\footnotesize 
(a) Redshift and (b) rest-frame 2500~\AA\ luminosity density distributions 
for the 206 SDSS RQQs in \rosat, \chandra, and \xmm\ 
observations. The shaded area indicates X-ray detections, 
the thin continuous line indicates the whole sample, and the thick line shows the RQQs detected by the RASS. 
\label{fig4}}
%\centerline{}
\end{figure*}
%%%%%%%%%%%%%%%%%%
%%%	END of FIG.4
%%%%%%%%%%%%%%%%%%%%%%%%%%%%%%%%%%%%%%%%%%%%%%%%%%%%%%%%%%%%%%%%%%%%%%%%%%%
%

\section{Broad-band properties of SDSS quasars}

One of the main goals of this paper is to study the broad-band properties 
of SDSS RQQs, parameterized by the optical-to-X-ray spectral index 
\hbox{
\aox=$\frac{\log(f_{2~\rm keV}/f_{2500~\mbox{\scriptsize\AA}})}{\log(\nu_{2~\rm keV}/\nu_{2500~\mbox{\scriptsize\AA}})}$
}, 
%=0.384\ $\log(L_{2~\rm keV}/L_{2500~\mbox{\scriptsize\AA}})$}, 
as a function of rest-frame UV luminosity density and redshift. 
Working with non-simultaneous 
UV and X-ray data will have some effects on the derived individual \aox\ values, but since our samples 
are fairly large we do not expect significant biases when the samples are considered as a whole. 
As most of the papers cited in $\S$1 
have focused on the relationship 
between rest-frame 2500~\AA\ luminosity density (hereafter referred to as \loo=\lo)
and rest-frame 2~keV luminosity density (hereafter referred to as \lxx=\lx), 
we will address this issue as well. 
The choice of using luminosity densities instead of flux densities is based on several arguments, 
as stated in Kembhavi, Feigelson, \& Singh (1986), and generally provides more robust results. 
However, this choice introduces a possible bias into the analysis, as luminosity is 
strongly correlated with redshift in flux-limited samples. 
It is therefore crucial to estimate the influence of this effect on the correlations 
in order to draw reliable conclusions about true physical relationships. 
In this paper, this has been achieved with the partial correlation analysis technique 
developed for use with censored data (Akritas \& Siebert 1996).

\subsection{The technique: Partial correlation and regression analysis}

A partial correlation analysis investigates the correlation of two variables 
while controlling for (holding constant) a third or additional variables. 
Partial correlation analysis requires meeting all of the 
assumptions of Pearsonian correlation, e.g., the linearity of the relationships 
(including that between the original variables and the control variable). 
The method used in this paper is that described by Akritas \& Siebert (1996); 
this allows one to apply partial correlation to censored data and to assign a significance 
level to the resulting correlation coefficient based on the Kendall $\tau$-statistic (Kendall 1970). 
When a significant (at the $\gax3\sigma$ confidence level) correlation is found according to 
the partial correlation analysis, its basic parameters are quantified via regression analysis, 
taking into account upper limits. 
For this purpose, we have used the {\sc ASURV} software package \hbox{Rev 1.2} (LaValley,
Isobe, \& Feigelson 1992), which implements the survival analysis methods presented 
in Feigelson \& Nelson (1985) and Isobe, Feigelson, \& Nelson (1986). 
We used the EM (Estimate and Maximize) regression algorithm (Dempster, Laird, \& Rubin 1977) 
and the Buckley-James (Buckley \& James 1979) regression method.

\subsection{The \rosat\ sample used in the \aox\ study}

Among the SDSS RQQs with \rosat\ coverage, 
in the \aox\ study we have used only the 128 RQQs 
covered by pointed PSPC and HRI observations. 
As shown in Fig.~3, the quasars detected by the RASS 
are strongly concentrated at bright X-ray fluxes, with most of the 
sources being brighter than \hbox{10$^{-13}$~\cgs} in the observed-frame 
\hbox{0.5--2~keV} band. Although the RASS covers 99.8\% of the sky, 
typical exposure times are \hbox{$\approx$~300--400~s} for most of 
the fields, which are not suitable for providing useful constraints 
on \aox\ for our sample. Including both the pointed and RASS data 
(detections and upper limits) in the \aox\ survival analyses below 
would likely lead to spurious results, as our sample would be dominated 
by weak RASS upper limits and would thus suffer from severe pattern 
censoring problems (see \S4.1.2.1 of Anderson 1985 and 
Appendix~A of Anderson \& Margon 1987). 

The \aox\ distribution of SDSS RQQs (excluding the RASS detections) is shown in Fig.~5 
(the shaded area indicates the \aox\ values for \hbox{X-ray} detected quasars).  
Taking into account upper limits using {\sc ASURV}, 
the average \aox\ is $-1.65\pm{0.02}$. 
%(the quoted errors represent the standard deviation of the mean). 
Note, however, that a dependence of \aox\ upon luminosity is likely present (see $\S$4.4). 

One object in our sample, SDSSp~J030422.39$+$002231.7 at $z=0.638$, is characterized by a notably 
steep \aox\ ($-2.01$), which suggests it could be a BALQSO or mini-BALQSO (e.g., Brandt et al. 2000). 
Its optical spectrum is quite blue and shows no obvious absorption near \ion{Mg}{2}. 
However, the [\ion{O}{3}]~5007\AA\ line is quite weak and the \ion{Fe}{2} emission-line blends are 
strong (G.~Richards 2002, private communication). 
%% From G. Richards
%This can be an argument in favor of it 
%being drawn from the parent sample of low-ionization BALQSOs (G.~Richards 2002, private communication). 
%
These properties have been seen from many low-ionization BALQSOs and other soft \hbox{X-ray} weak quasars 
(e.g., Boroson \& Meyers 1992; Turnshek et al. 1997; Laor \& Brandt 2002). 
%
%%%%%%%%%%%%%%%%%%%%%%%%%%%%%%%%%%%%%%%%%%%%%%%%%%%%%%%%%%%%%%%%%%%%%%%%%%%
%%%	FIGURE 5: aox distribution excluding the RASS objects
%%%%%%%%%%%%%%%%%%
%\begin{figure}
\figurenum{5}
\centerline{\includegraphics[angle=0,width=8.5cm]{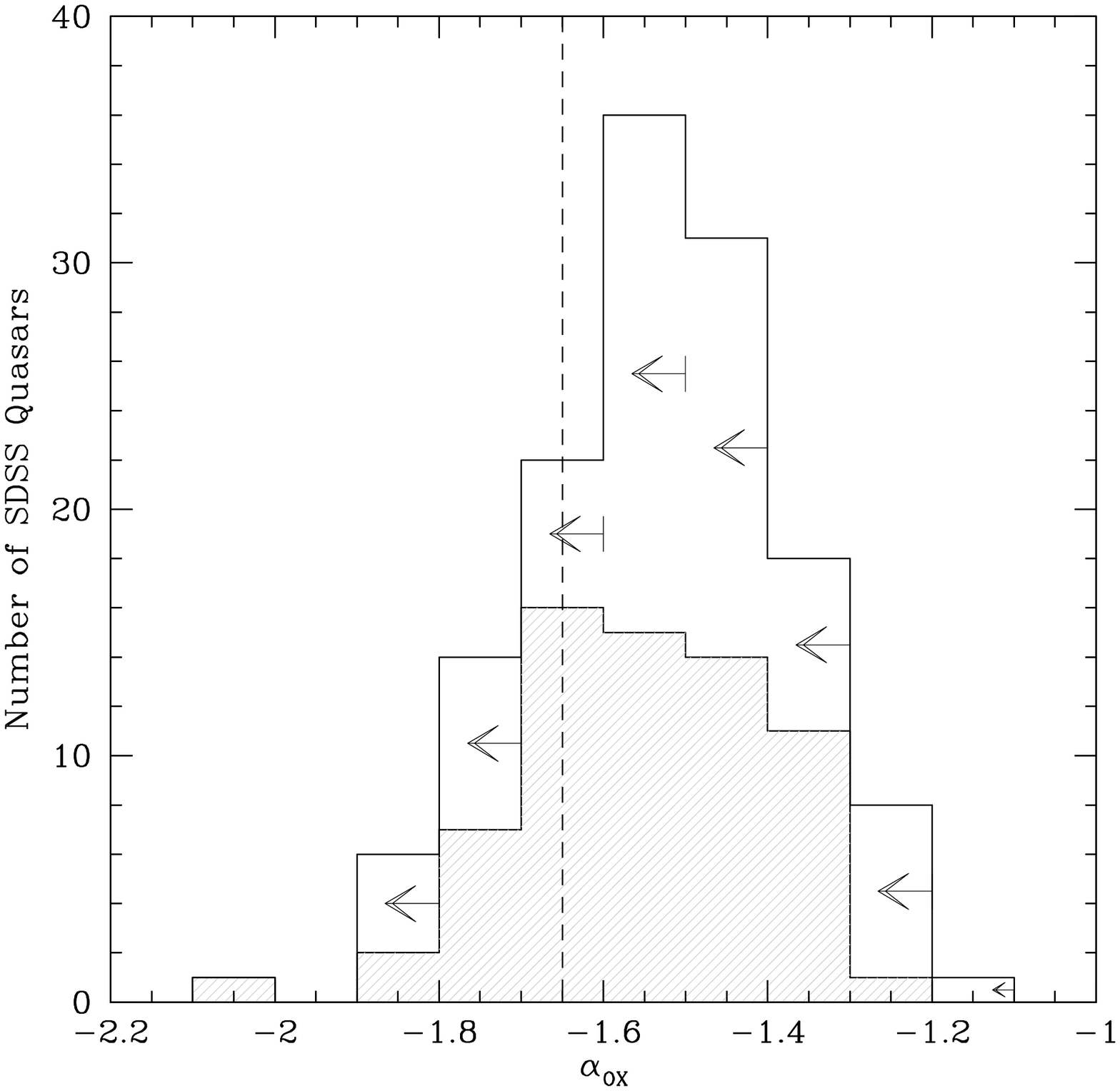}}
\figcaption{\footnotesize 
Histogram of the \aox\ distribution for SDSS RQQs with coverage in pointed X-ray observations. 
The shaded area indicates the quasars with 
X-ray detections, while the continuous line indicates those from the entire sample. 
Unshaded areas with arrows indicate \aox\ upper limits. 
The vertical dashed line at \aox=$-$1.65 indicates the average \aox\ calculated in $\S$4.2. 
\label{fig5}}
\centerline{}
%\end{figure}
%%%%%%%%%%%%%%%%%%
%%%	END of FIG.5
%%%%%%%%%%%%%%%%%%%%%%%%%%%%%%%%%%%%%%%%%%%%%%%%%%%%%%%%%%%%%%%%%%%%%%%%%%%

\subsection{The effect of unknown BALQSOs at $z\lax1.5$ on \aox\ measurements}

As mentioned in $\S$3.1.1, at $z\lax1.5$ it is not possible to identify BALQSOs among 
the quasars from SDSS spectra. However, their importance in average \aox\ determinations can be estimated statistically. 
Given the high fraction of X-ray undetected BALQSOs at $z>1.5$ (80\%) and the general evidence that BALQSOs 
have strongly depressed X-ray emission in the \hbox{0.5--2~keV} band (see $\S$1), we have randomly removed from the $z<1.5$ subsample 
(69 RQQs with 39 X-ray detections) 
a variable number (10--15\% of this subsample) of X-ray non-detections. 
Taking into account the upper limits using {\sc ASURV}, the average \aox\ at $z<1.5$ is 
\hbox{$\langle\alpha_{\rm ox}\rangle$ = $-$1.61$\pm{0.03}$}. 
By randomly excluding some of the \hbox{X-ray} non-detections 
as previously described, 
we obtain \hbox{$\langle\alpha_{\rm ox}\rangle$ = $-$1.58$\pm{0.03}$}, close to the previous value. 
It must be kept in mind that some BALQSOs could be detected at $z<1.5$; therefore 
the previous average value of $-$1.58 would represent a conservative estimate.

\subsection{\aox\ versus redshift and 2500~\AA\ luminosity density}

The \aox\ distributions as a function of redshift (Fig.~6a) and 2500~\AA\ luminosity 
density (Fig.~7a) are characterized by large scatter, which prevents one from easily identifying a trend. 
To shed light on this issue, we averaged the \aox\ values in redshift (Fig.~6b) 
and luminosity (Fig.~7b) bins taking into account the upper limits. 
The flat \aox\ value found in the redshift range $3<z<4$ is likely to be due to the small number of objects 
(eight RQQs, five of which are non-detections). 
A trend of decreasing \aox\ as a function of redshift and luminosity 
can be seen in both distributions, although it is not possible to determine which anti-correlation is 
the most important one or to quantify the significance without applying partial correlation analysis. 
It must be kept in mind that a certain degree of anti-correlation is expected between \aox\ and \lo, 
given the presence of the UV luminosity 
%
%%%%%%%%%%%%%%%%%%%%%%%%%%%%%%%%%%%%%%%%%%%%%%%%%%%%%%%%%%%%%%%%%%%%%%%%%%%
%%%	FIGURE 6: aox vs. Redshift: unbinned and binned data [NO RASS]
%%%%%%%%%%%%%%%%%%
%\begin{figure}[!h]
%\vglue -0.1cm
\figurenum{6}
\resizebox{8.5cm}{!}{\includegraphics{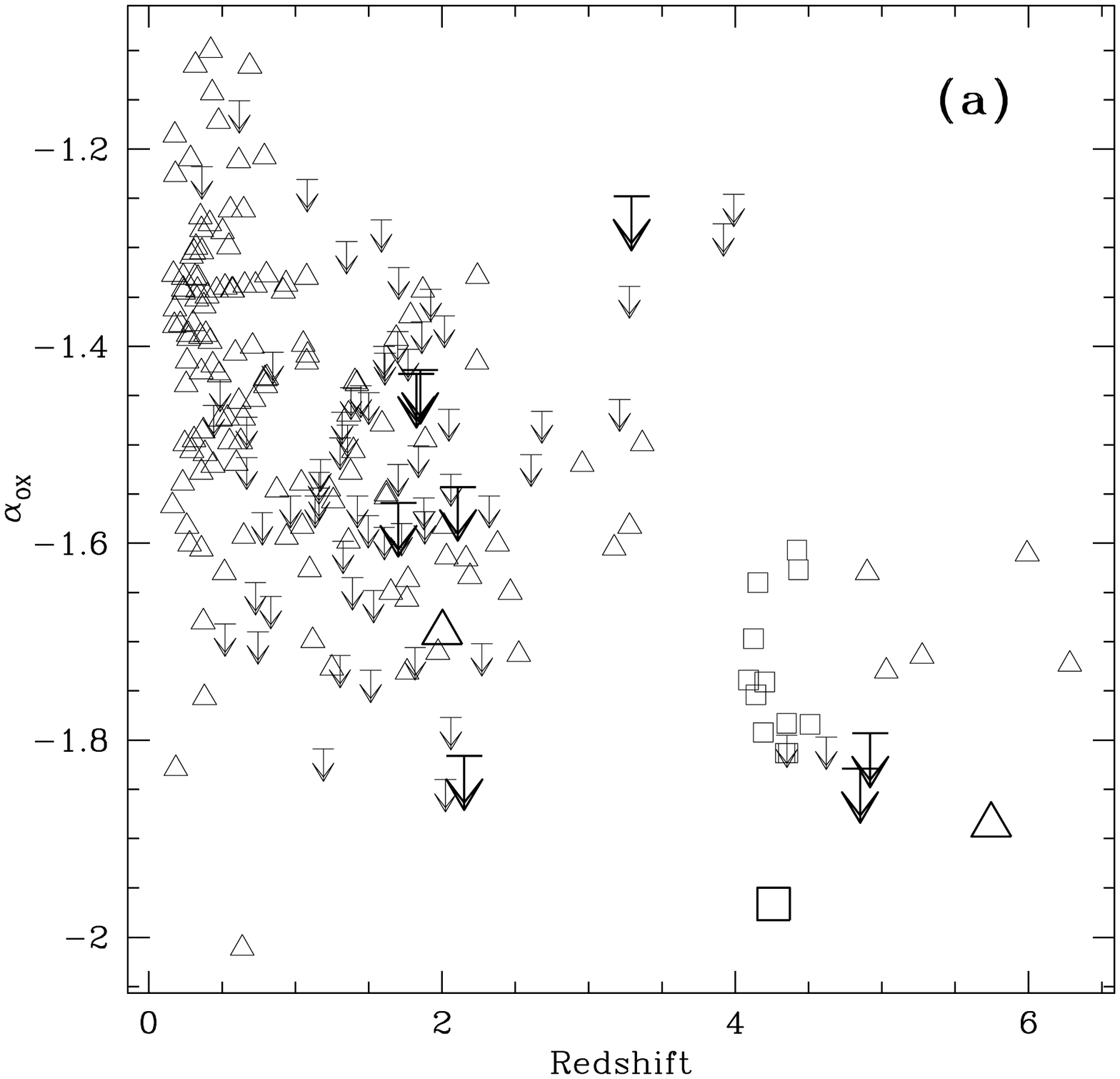}}
%\vglue-0.15cm
%\hspace{1cm}
\resizebox{8.5cm}{!}{\includegraphics{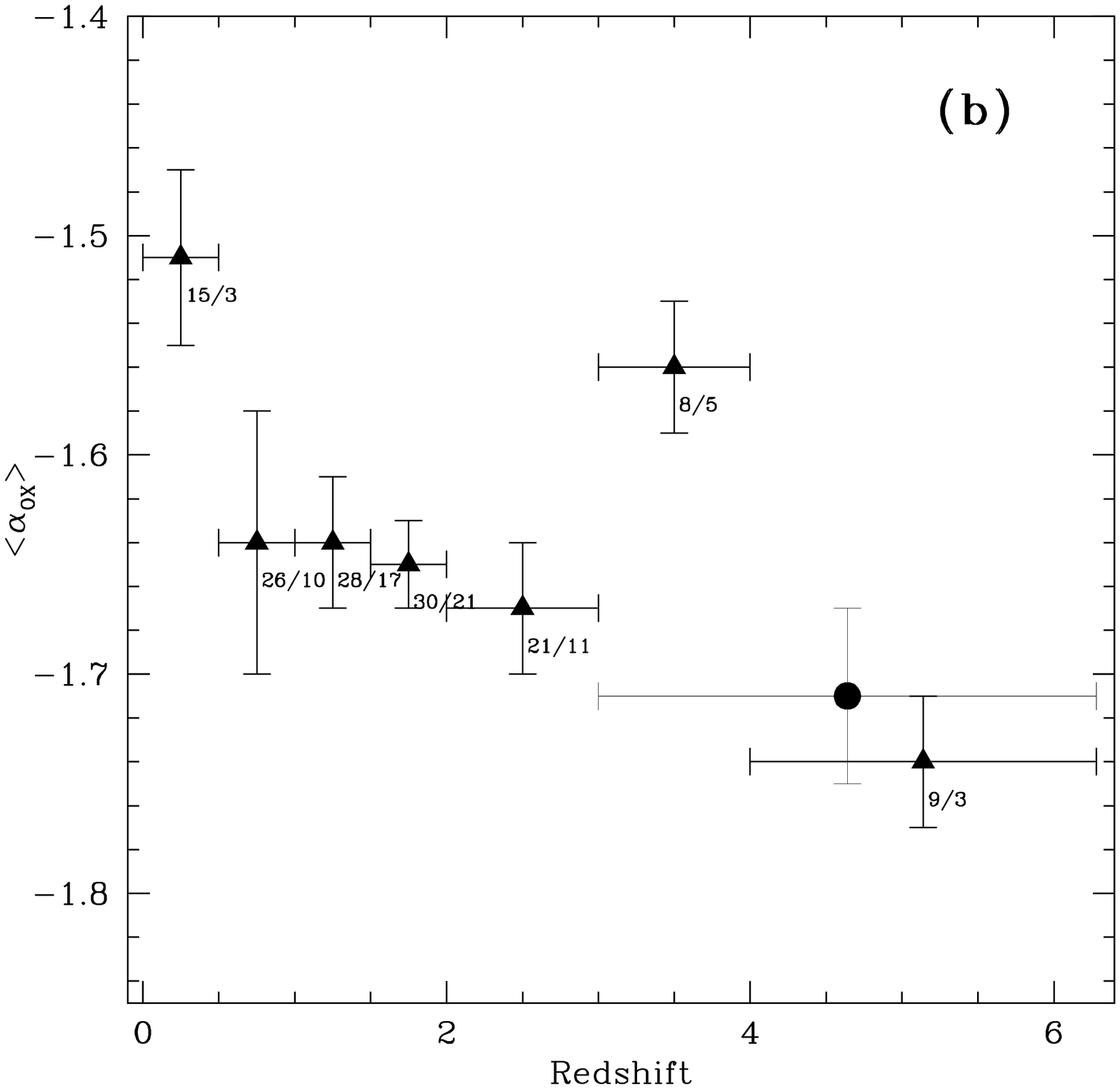}}
\figcaption{\footnotesize 
(a) \aox\ distribution of RQQs as a function of redshift. 
SDSS RQQs are shown as triangles (X-ray detections) and downward-pointing arrows (X-ray upper limits). 
The open squares indicate the PSS quasars (12 X-ray detected RQQs and one upper limit). 
Larger and thicker symbols indicate the BALQSOs. 
(b) Average \aox\ indices as a function of redshift for the SDSS RQQs only. The \aox\ values have been averaged using {\sc ASURV} 
(LaValley et al. 1992). The two numbers associated with each data point (triangles) indicate 
the total number of quasars and the number of upper limits in each redshift interval 
(whose range is shown by the $x$-axis bar), respectively. 
The $y$-axis error bars represent the standard deviation of the mean (as determined with {\sc ASURV}). 
To improve the statistics, the average \aox\ value obtained using all quasars at $z>3$ is also 
displayed (filled circle). 
\label{fig6}}
\centerline{}
\centerline{}
%\end{figure}
%%%%%%%%%%%%%%%%%%
%%%	END of FIG.6
%%%%%%%%%%%%%%%%%%%%%%%%%%%%%%%%%%%%%%%%%%%%%%%%%%%%%%%%%%%%%%%%%%%%%%%%%%%
\noindent term in the \aox\ formula (see $\S$4.5 for further discussion). 

The measured values of \aox\ show no strong dependence upon 
redshift (2.1$\sigma$; 1.7$\sigma$ when the BALQSOs are removed from the analysis; see Table~3) from the 
partial correlation analysis. 
We also searched for possible \aox--redshift dependencies in individual redshift/luminosity bins, 
but found none. 
The inclusion of the 13 PSS quasars (see $\S$3.2; open squares in Fig.~6a) provides a 
slightly, but not sufficiently, better significance of the 
above anti-correlation (2.8$\sigma$; 2.3$\sigma$ when the BALQSOs are removed). 

%%%%%%%%%%%%%%%%%%%%%%%%%%%%%%%%%%%%%%%%%%%%%%%%%%%%%%%%%%%%%%%%%%%%%%%%%%%
%%%	TABLE 3: Correlations and their significances from partial correlation analysis
%%%%%%%%%%%%%%%%%%
%%\begin{table*}[!t]
\footnotesize
%\caption{Correlations and their significances from partial correlation analysis}
\begin{center}
{\sc TABLE 3 \\ Correlations and their significances from partial correlation analysis}
\vskip 4pt
\begin{tabular}{lccc}
\hline
\hline
  & \aox--$z$  &  & \aox--\loo \\
\hline
SDSS only, with BALQSOs       & 2.1$\sigma$ & & 3.7$\sigma$ \\
SDSS only, without BALQSOs    & 1.7$\sigma$ & & 3.4$\sigma$ \\
SDSS and PSS, with BALQSOs    & 2.8$\sigma$ & & 3.8$\sigma$ \\
SDSS and PSS, without BALQSOs & 2.3$\sigma$ & & 3.9$\sigma$ \\
\hline
\end{tabular}
%%\vskip 2pt
\end{center}
\setcounter{table}{3}
\label{tab3}
\centerline{}
\centerline{}
%%\end{table*}
\normalsize
%%%%%%%%%%%%%%%%%%%%%
%%%	End of Table 3
%%%%%%%%%%%%%%%%%%%%%%%%%%%%%%%%%%%%%%%%%%%%%%%%%%%%%%%%%%%%%%%%%%%%%%%%%%%
%

Applying the partial correlation analysis, we find that 
the \aox--\loo\ anti-correlation 
(hereafter parameterized by the equation $\alpha_{\rm ox}=A_{\rm l}$\ \loo+$B_{\rm l}$) 
is significant at the 3.7$\sigma$ level (3.4$\sigma$ level) 
when the BALQSOs are included (excluded) in the analysis (see Table~3). 
According to the EM regression, the parameters of this relationship are 
\begin{equation}
A_{\rm l}=-0.11\pm{0.02} \hspace{0.3cm}
B_{\rm l}=1.85\pm{0.69} \hspace{0.2cm} {\rm (including~~BALQSOs)}
\end{equation}
\begin{equation}
A_{\rm l}=-0.10\pm{0.02} \hspace{0.3cm}
B_{\rm l}=1.32\pm{0.70} \hspace{0.2cm} {\rm (excluding~~BALQSOs)}.
\end{equation}
The first parameterization is shown in Fig.~7a as a dashed line. 
Similar values are found using the Buckley-James regression method. 
Using the same approach described in $\S$4.3, we evaluated the impact of 
possible unknown BALQSOs in the partial correlation analysis 
by randomly removing $\approx$~10\% of the \hbox{X-ray} non-detections at $z<1.5$. 
% 70 from pointed PSPC and HRI observations at z<=1.5. 
The \aox--\loo\ anti-correlation is slightly more significant after this procedure.   
The \aox--\loo\ anti-correlation becomes slightly more significant (3.8$\sigma$) 
after the inclusion of the PSS 
quasars in the analysis (open squares in Fig.~7a), with similar best-fit parameters. 
Our finding above that \aox\ depends primarily on \loo\ is consistent with the fact that the 
\aox\ dependence upon \loo\ in Fig.~7b appears somewhat stronger than the \aox\ dependence upon 
redshift in Fig.~6b. 
Given the dependence of \aox\ upon \loo, 
the assumption of the same \aox\ value regardless of the UV luminosity density of a quasar 
can imply up to a factor of $\approx$~3 error in estimating its soft X-ray flux. 

It is worth noting that \hbox{$\approx$~5--10\%} of \hbox{$z\approx$~4--6} quasars 
may experience a factor $\gax2$ magnification 
due to gravitational lensing (e.g., Wyithe \& Loeb 2002; Comerford, Haiman, \& Schaye 2002), assuming the 
SDSS luminosity function reported by Fan et al. (2001b) for \hbox{$M_{\rm B}<-26$} SDSS quasars 
[$\phi(L) \propto L^{\beta}$, with $\beta\approx-2.5$]. 
For steeper luminosity functions, the fraction of high-redshift lensed quasars may increase to \hbox{$\approx$~30\%} 
(e.g., Wyithe \& Loeb 2002). Such an effect will confuse studies of the dependence of \aox\ (which should not be 
changed by lensing, unless the UV and \hbox{X-ray} luminosities 
are magnified differently) upon luminosity. 
Imaging of known $z>4$ quasars with \hst\ will 
soon constrain the fraction of gravitationally lensed high-redshift quasars (M. Strauss 2002, private communication). 
We have roughly estimated the importance of such an effect by randomly reducing by a factor $\approx$~2--5 the 
UV luminosity densities of $\approx$~10\% (i.e., \hbox{2--3} objects) of the $z>4$ quasars, including the PSS sample. 
The partial correlation results from this exercise do not change significantly from the previous analysis; 
the \aox--\loo\ relation is still present. 
%
%%%%%%%%%%%%%%%%%%%%%%%%%%%%%%%%%%%%%%%%%%%%%%%%%%%%%%%%%%%%%%%%%%%%%%%%%%%
%%%	FIGURE 7: aox vs. 2500 A Luminosity: unbinned + binned data [NO RASS]
%%%%%%%%%%%%%%%%%%
%\begin{figure}
\figurenum{7}
\resizebox{8.5cm}{!}{\includegraphics{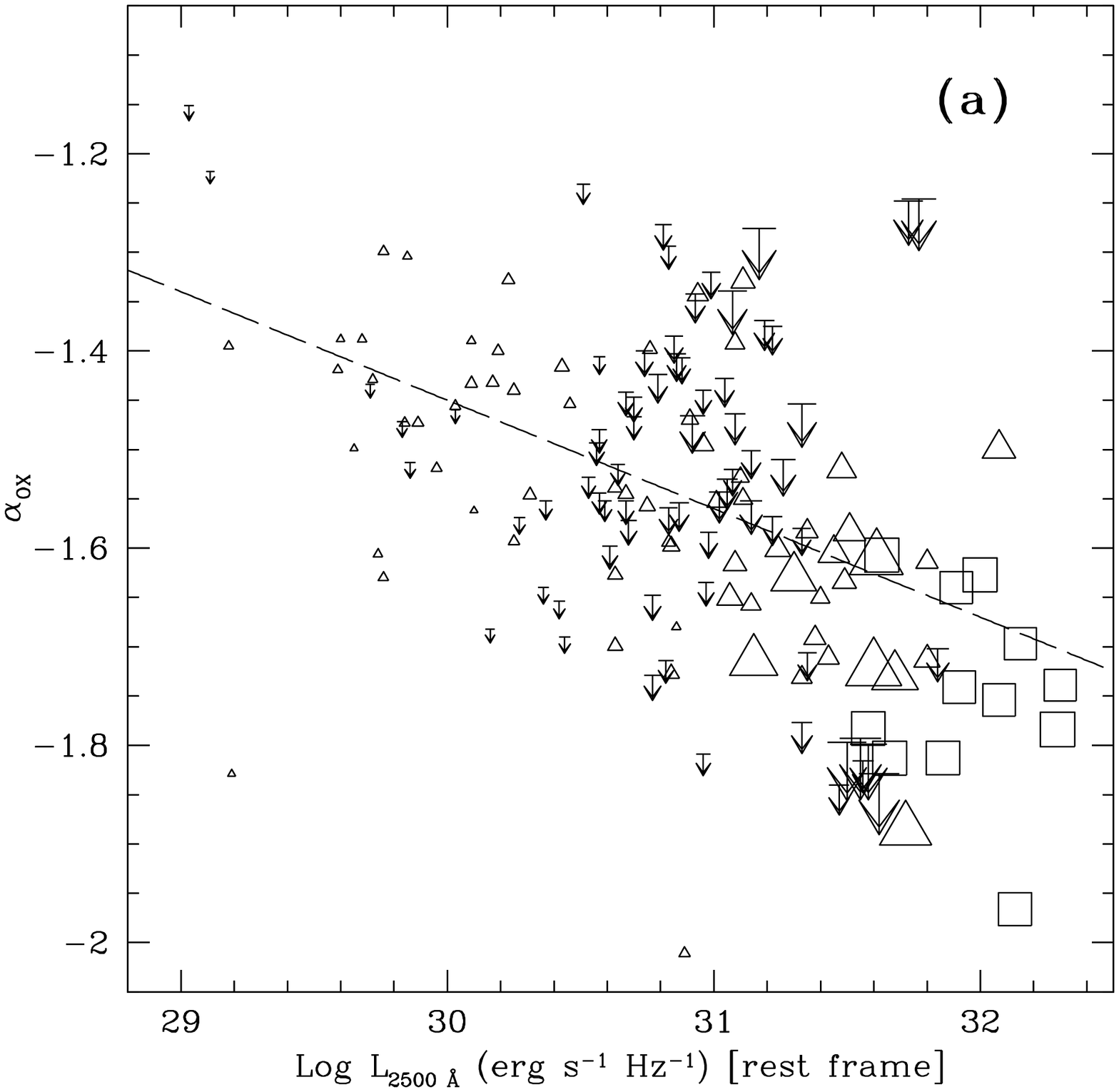}}
\resizebox{8.5cm}{!}{\includegraphics{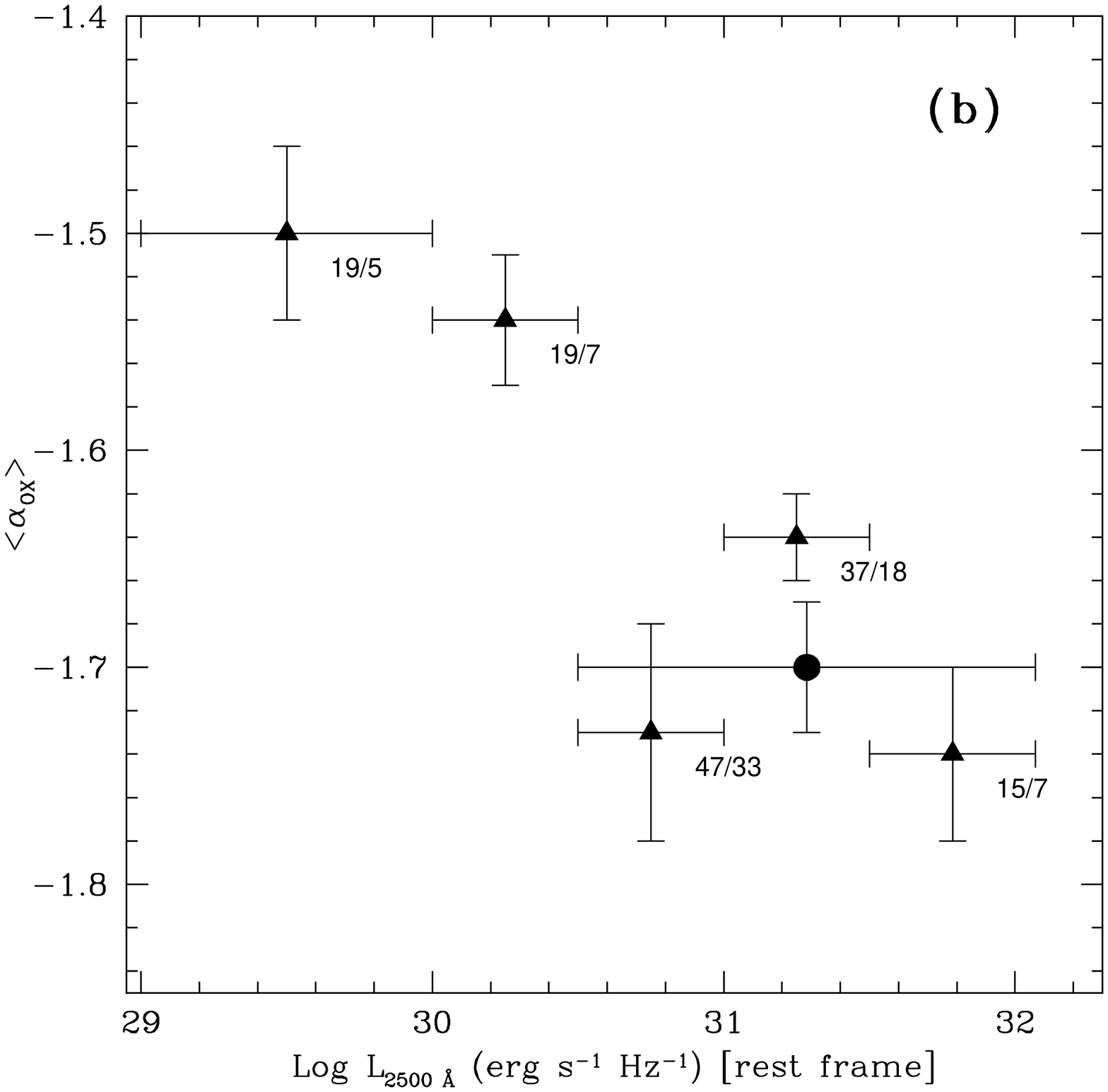}}
\figcaption{\footnotesize 
(a) \aox\ distribution as a function of rest-frame 2500~\AA\ luminosity. 
The sizes of the symbols (triangles for the SDSS X-ray detected RQQs and downward-pointing arrows for the 
upper limits) increase as a function of redshift. The open squares indicate the PSS quasars (12 X-ray detected RQQs and one upper limit). 
The dashed line indicates the relationship obtained with the pure SDSS sample using {\sc ASURV} 
and including the BALQSOs (equation~1). 
% but excluding the PSS quasars. 
%
(b) 
Average \aox\ indices as a function of rest-frame 2500~\AA\ luminosity for the SDSS RQQs only. The \aox\ values have been averaged using 
{\sc ASURV} (LaValley et al. 1992). The two numbers associated with each data point indicate 
the total number of quasars and the number of upper limits in each luminosity interval 
(whose range is shown by the $x$-axis bar), respectively. 
The $y$-axis error bars represent the standard deviation of the mean. 
To improve the statistics, the average \aox\ value obtained using all quasars at \loo$>$30.5~\lumh\ is also 
displayed (filled circle). 
\label{fig7}}
\centerline{}
\centerline{}
%\end{figure}
%%%%%%%%%%%%%%%%%%
%%%	END of FIG.7
%%%%%%%%%%%%%%%%%%%%%%%%%%%%%%%%%%%%%%%%%%%%%%%%%%%%%%%%%%%%%%%%%%%%%%%%%%%

Finally, we have assessed if our main results above are sensitive to our chosen cosmology by repeating 
our analyses with a $H_0$=50 km s$^{-1}$ Mpc$^{-1}$ and $q_0$=0.0 cosmology (see $\S$1). 
Our main results appear qualitatively unchanged for this choice of cosmology.

\subsection{\lxx\ versus \loo}

Previous studies of optically selected quasars have found 
a strong correlation between optical and X-ray luminosity densities (e.g., Zamorani et al. 1981), 
parameterized by \lxx\ $\propto$ \loo$^{\beta}$, 
with $\beta$ typically varying from $\approx$~0.7 (e.g., Pickering et al. 1994; Wilkes et al. 1994) to $\approx$~0.8 
(e.g., Avni \& Tananbaum 1986). 
Most of these studies were based on \einstein\ data, and the results were 
obtained using the Detection and Bounds method described by Avni et al. (1980), i.e., taking 
into account both detections and upper limits. 
Using a different approach, stacking analysis, Green et al. (1995) obtained a similar result 
($\beta\approx$~0.86) for a sample of Large Bright Quasar Survey (LBQS; Hewett, Foltz, \& Chaffee 1995) 
quasars observed in the RASS. 
On the other hand, a study by La Franca et al. (1995), based on \einstein\ data, 
found that the slope for the ultraviolet--X-ray relationship 
was consistent with unity using the modified orthogonal distance regression method 
(Fasano \& Vio 1988), which accounts for 
errors in both variables and intrinsic scatter in the data. 
However, as pointed out by Yuan et al. (1998a), the significant dispersion in the data and, possibly, their 
inhomogeneous object selection, could induce a steeper correlation slope. 

Finding a non-linear \lxx--\loo\ correlation (with a slope less than unity) for our SDSS sample of RQQs 
would support our previous results (see $\S$4.4) 
in a totally unbiased way. While a certain degree of anti-correlation is expected between \aox\ and 
\loo\ given the presence of the UV luminosity term in the \aox\ formula, no correlation is mathematically expected {\it a priori} 
between \lxx\ and \loo. 
With the present work we confirm previous results about the existence of a 
strong correlation between \lxx\ and \loo, with an extension to the highest redshifts/luminosities 
given the inclusion of $z>4$ quasars recently observed by \chandra\ and \xmm. 
We used the generalized Kendall $\tau$-statistic (Brown, Hollander, \& Korwar 1974) 
to evaluate the significance of the result, and the 
EM algorithm (Dempster et al. 1977) and the Buckley-James method (Buckley \& James 1979) 
to derive the parameters of the relationship. 
We find a significant correlation (see Fig.~8), parameterized by 
\begin{equation}
{\it l_x}\propto{\it l_{uv}}^{0.71\pm{0.06}} \hspace{0.2cm} {\rm (including~~BALQSOs; 7.8\sigma~significance~level)}
\end{equation}
and 
\begin{equation}
{\it l_x}\propto{\it l_{uv}}^{0.75\pm{0.06}} \hspace{0.2cm} {\rm (excluding~~BALQSOs; 7.9\sigma~significance~level)}
\end{equation}
The latter correlation is shown in Fig.~8 (dashed line). For comparison purposes, 
the dotted line shows a correlation with $\beta=1$ constrained to pass through the mean \lxx\ and \loo. 

The inclusion of the PSS quasars at $z>4$ has the effect of increasing the significance of this 
correlation up to 9.3$\sigma$ (both with and without the BALQSOs), and the basic parameters of the correlation remain  
unchanged. 
The significance of this correlation over a large redshift and luminosity range 
suggests that it is a universal property of optically selected (and also X-ray selected; 
e.g., La Franca et al. 1995) quasars. 
%
%%%%%%%%%%%%%%%%%%%%%%%%%%%%%%%%%%%%%%%%%%%%%%%%%%%%%%%%%%%%%%%%%%%%%%%%%%%
%%%	FIGURE 8: lo versus lx for the whole sample (209 objects) with the 
%%%               main correlation lines overlaid
%%%%%%%%%%%%%%%%%%
%\begin{figure}
\figurenum{8}
\centerline{\includegraphics[angle=0,width=8.5cm]{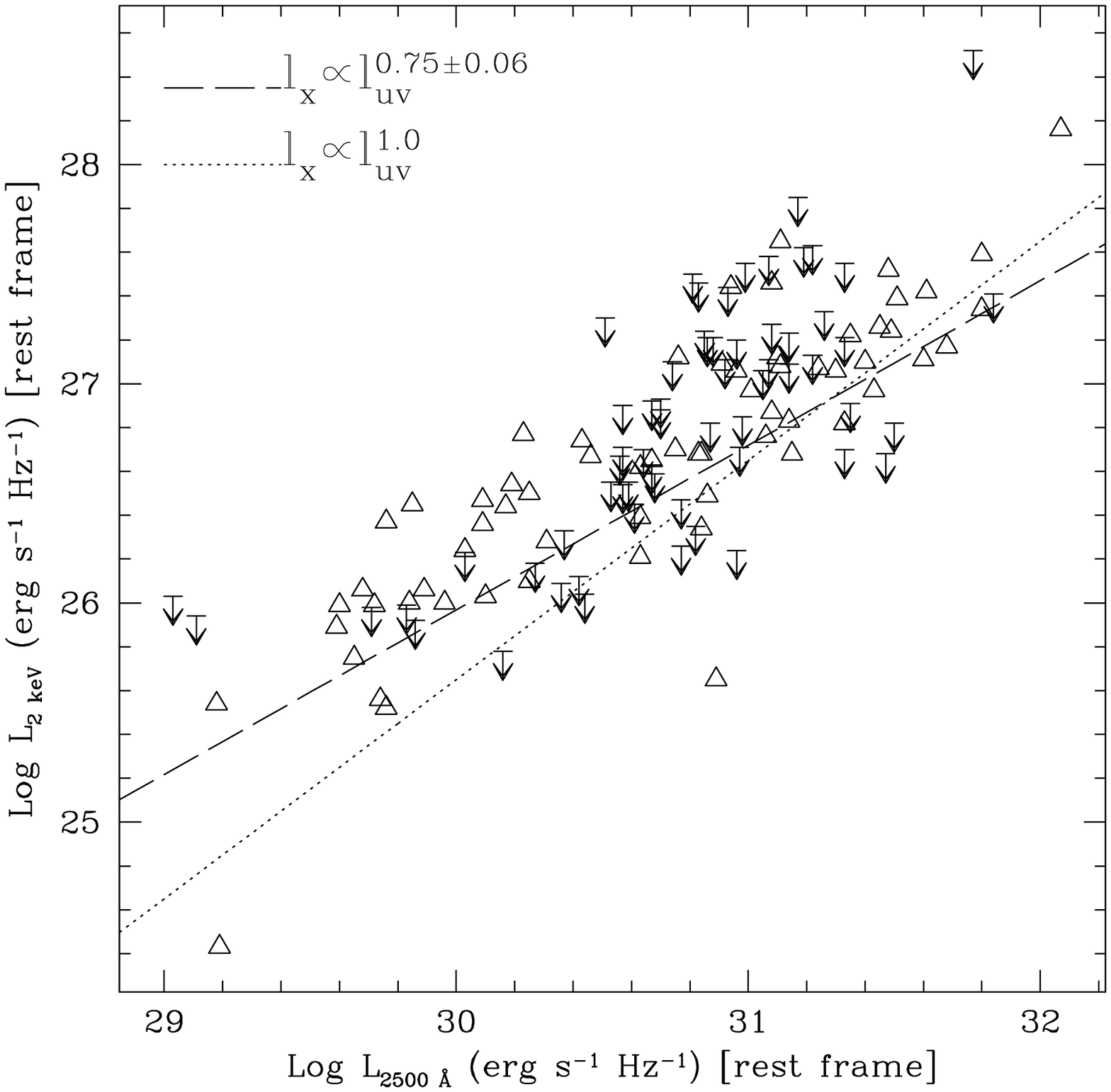}}
\figcaption{\footnotesize 
\lxx\ versus \loo. Triangles indicate SDSS RQQs observed in pointed \hbox{X-ray} observations (excluding BALQSOs). 
%The symbols are the same as in previous figures. 
The dashed line indicates the best-fit relationship obtained 
when BALQSOs are excluded from the analysis. 
For comparison purposes, the dotted line shows a correlation with $\beta=1$, 
constrained to pass through the average \lxx\ and \loo.
\label{fig8}}
\centerline{}
\centerline{}
%\end{figure}
%%%%%%%%%%%%%%%%%%
%%%	END of FIG.8
%%%%%%%%%%%%%%%%%%%%%%%%%%%%%%%%%%%%%%%%%%%%%%%%%%%%%%%%%%%%%%%%%%%%%%%%%%%

We have investigated if $\beta$ has any redshift dependence. 
Using the RQQs at $z<1.5$ ($z>1.5$) we find $\beta=0.61\pm{0.11}$ 
($\beta=0.90\pm{0.19}$ excluding BALQSOs and $\beta=0.79\pm{0.20}$ 
including BALQSOs); the division at $z=1.5$ has been chosen to give approximately the same 
number of objects in each redshift interval. Therefore, we do not detect any highly significant variation 
of $\beta$ with redshift. 

Using the definition of \aox, the equation $\alpha_{\rm ox}=A_{\rm l}$\ \loo+$B_{\rm l}$ (see $\S$4.4), 
and \lxx\ $\propto$ \loo$^{\beta}$, 
we can derive $A_{\rm l}$ as a function of $\beta$ and some constant terms. 
Using the range of \loo\ for our SDSS sample and the constant terms reported in equations (2) and (4), 
we derive $A_{\rm l}\approx$~$-$0.10. This is entirely consistent with the value reported in equation (2).

\section{Discussion}

\subsection{Results from the SDSS RQQ sample}

We have presented the \hbox{X-ray} properties of the color-selected SDSS RQQs, mainly taken from the EDR, 
using \rosat, \chandra, and \xmm\ data. 
In the 0.16--6.28 redshift range, 
we detect 67/137 SDSS RQQs covered by pointed \hbox {X-ray} observations and an additional 69 SDSS RQQs 
observed by the RASS. 
We have used the unbiased sample of 137 RQQs with pointed \hbox{X-ray} observations 
to investigate the dependence of the \aox\ index 
%%
%Using an unbiased sample (i.e., excluding the 69 quasars with RASS detections), 
%we have investigated the dependence of the \aox\ index 
%%
upon rest-frame 2500~\AA\ luminosity and redshift using partial correlation analysis applied to censored \hbox{X-ray} data. 
The principal results of our study are the following: 
\begin{itemize}

\item 
The optical-to-X-ray spectral slope, \aox, is anti-correlated with \loo\ at the 3.4--3.7$\sigma$ significance level, 
although the scatter in this anti-correlation is large (see $\S$4.4). 
The inclusion of 13 $z>4$, high-luminosity PSS quasars increases slightly the significance of the 
\aox--\loo\ anti-correlation to 3.8--3.9$\sigma$. 
There is no significant correlation between \aox\ and redshift, even when the sample of PSS quasars is included. 
These results suggest that the mechanism driving quasar broad-band emission 
(i.e., accretion onto a super-massive black hole) is similar in the local and the early Universe. 
They also indicate that, although it is likely that high-redshift quasars are radiating at higher fractions of their Eddington 
luminosity (e.g., Kauffmann \& Haehnelt 2000), no unusual phenomena such as 
accretion-disk instabilities (e.g., Lightman \& Eardley 1974; Shakura \& Sunyaev 1976) or 
``trapping radius'' effects (e.g., Begelman 1978; Rees 1978) seem to be dominant. 
It appears that already at $z\approx$~6 
quasars are almost completely built-up systems, even though 
the time available was short at those epochs ($\simlt1$~Gyr in our cosmology). 

\item
We confirm the \lxx--\loo\ correlation as being an intrinsic property for optically selected RQQs. 
This correlation can be parameterized by \lxx$\propto$\loo$^{0.75\pm{0.06}}$, 
in good agreement with most previous work. 

\end{itemize}

\subsection{Comparison with earlier work}

The \aox\ dependence upon luminosity and redshift has been studied for many years. 
However, most earlier studies, although generally providing results consistent with ours (i.e., 
\aox\ depends mainly upon \loo), were limited by heterogeneous selection criteria of the samples (e.g., 
Avni et al. 1995) or limited coverage in redshift/luminosity. 
As pointed out by Yuan, Siebert, \& Brinkmann (1998b) using Monte-Carlo simulations, 
a spurious anti-correlation between \aox\ and UV luminosity density 
can emerge even for a population with an intrinsically constant \aox, 
provided that the dispersion of the UV luminosities 
%deviating from the average SED 
is similar to or larger than that of the X-ray luminosities. 
Analysis of the dispersions of \loo\ and \lxx\ for 
the SDSS RQQs used in this work indicates that the \aox--\loo\ 
anti-correlation is likely to be real. 

The lack of sensitive radio surveys (e.g., FIRST, Becker et al. 1995; NVSS, Condon et al. 1998) 
prevented the early studies of quasars from clearly discriminating between the radio-quiet and 
radio-loud populations (except for the sources detected at high radio flux densities); these two populations 
have been known to have different emission properties in the X-ray band 
since \einstein\ (e.g., Zamorani et al. 1981; Wilkes \& Elvis 1987). 
In this regard, the sensitive radio coverage for all of the quasars used in our study 
allowed us to select a sample of ``pure'' RQQs without contamination by radio-loud objects. 

The only results indicating that \aox\ depends upon redshift have been presented by 
Yuan et al. (1998a) and Bechtold et al. (2002). 
Yuan et al. (1998a), using a large database of RQQs seen in the RASS and \rosat\ pointed observations, 
found that \aox\ depends on 2500~\AA\ luminosity at \hbox{\loo$>30.5$~\lumh\ }and slightly on redshift at $z<0.5$. 
We did not find the latter dependence, 
but the number of SDSS RQQs with \hbox{X-ray} coverage at $z<0.5$ is small (15 RQQs). 
Bechtold et al. (2002) reported a significant \aox--redshift 
anti-correlation for a large sample of quasars observed by \rosat\ (data taken from 
Yuan et al. 1998a) and \chandra\ over the same redshift range as our study. 
The heterogeneous selection and construction criteria of the Bechtold et al. (2002) sample, coupled with the 
lack of a partial correlation analysis, might lead to different results than ours.

\subsection{Future possibilities and work}

Studies using a larger number of objects and, hopefully, a higher fraction of \hbox{X-ray} detections 
are required to address all the issues related to \aox\ dependences upon luminosity or redshift. 
A larger number of SDSS quasars will soon become available to the scientific community 
with the next data release, which will provide a factor of $\approx$~5 increase in the number of SDSS optically selected objects. 
From an \hbox{X-ray} perspective, most of these quasars 
are likely to be detected easily with snapshot observations by modern \hbox{X-ray} satellites. 
Furthermore, given the large field-of-view of the EPIC instruments onboard 
\xmm\ ($\approx$~15\arcmin\ radius), we expect a sizeable 
number of serendipitously observed SDSS RQQs in the next few years.

\acknowledgments

We gratefully acknowledge the financial support of \chandra\ X-ray Center 
grant DD1-2012X (CV, WNB, DPS), NASA LTSA grant NAG5-8107 (CV, WNB), 
and NSF grant AST99-00703 (DPS). 
We thank M.~G. Akritas, D.~M. Alexander, S.~F. Anderson, 
F.~E. Bauer, A. Comastri, E.~D. Feigelson, L. Pozzetti, G. Richards, and G. Zamorani 
for interesting and thoughful discussions, and J. Englhauser for help with the RASS archive. 
CV also acknowledges partial support from the Italian Space Agency, 
under the contract ASI I/R/113/01, 
and from the Ministry for University and Research (MURST) 
under grant \hbox{Cofin-00-02-36}. 

Funding for the creation and distribution of the SDSS Archive has been provided by 
the Alfred P. Sloan Foundation, the Participating Institutions, the National Aeronautics 
and Space Administration, the National Science Foundation, the U.S. Department of Energy, 
the Japanese Monbukagakusho, and the Max Planck Society. 
The SDSS Web site is http://www.sdss.org/. 
The SDSS is managed by the Astrophysical Research Consortium (ARC) for the Participating Institutions. 
The Participating Institutions are the University of Chicago, Fermilab, the Institute for Advanced Study, 
the Japan Participation Group, the Johns Hopkins University, Los Alamos National Laboratory, 
the Max-Planck-Institute for Astronomy (MPIA), the Max-Planck-Institute for Astrophysics (MPA), 
New Mexico State University, the University of Pittsburgh, Princeton University, the United States Naval Observatory, 
and the University of Washington.

%\clearpage

\newpage

%%%%%%%%%%%%%%%%%%%%%%%%%%%%%%%%%%%%%%%%%%
% aox BIG TABLE
%%%%%%%%%%%%%%%
\begin{deluxetable}{lccccccrrrrrrcrrr}
\setcounter{table}{1}
\rotate
\tablecolumns{17}
\tabletypesize{\tiny}
\tablewidth{0pt}
\tablecaption{Properties of the SDSS quasars with X-ray Observations}
\tablehead{
\colhead{SDSS Object} & \colhead{Alt. Name\tablenotemark{a}} & \colhead{$z$} & \colhead{$AB_{1450(1+z)}$} & \colhead{$M_B$} & \colhead{$f_{2500~\AA}$\tablenotemark{b}} & 
\colhead{$\log L_{2500~\AA}$\tablenotemark{c}} & \colhead{$f_{\rm 0.5-2~keV}$\tablenotemark{d}} & \colhead{$f_{\rm 2~keV}$\tablenotemark{b}} & 
\colhead{$\log L_{\rm 2~keV}$\tablenotemark{c}} & \colhead{$\alpha_{\rm ox}$} & \colhead{$f_{\rm 1.4~GHz}$\tablenotemark{e}} & 
\colhead{$R$\tablenotemark{f}} & \colhead{Instr.\tablenotemark{g}} & \colhead{Exp.\tablenotemark{h}} & \colhead{Off-Axis\tablenotemark{i}} & \colhead{Obs. ID} 
}
\startdata
000710.01$+$005329.0     & LBQS 0004+0036        &  0.316  &  17.67  &  $-$23.95  &  5.07  &  30.10  &	  8.85    &     1.74   &	26.63  &   $-$1.330  &     1.55  &      0.93  &	 R  &    396  &	       &  rs931701n00  \\
000834.71$+$003156.1     & LBQS 0006+0015        &  0.263  &  18.26  &  $-$22.70  &  2.94  &  29.70  &	  3.21    &     0.604  &	26.01  &   $-$1.415  &  $<0.44$  &   $<0.56$  &	 R  &    395  &	       &  rs931701n00  \\
001257.25$+$011527.3     &                       &  0.504  &  18.90  &  $-$23.31  &  1.59  &  30.02  &	  3.21    &     0.721  &	26.67  &   $-$1.284  &  $<0.36$  &   $<1.14$  &	 R  &    384  &	       &  rs931701n00  \\
002209.95$+$001629.3     & LBQS 0019-0000        &  0.575  &  18.58  &  $-$24.05  &  2.22  &  30.28  &	  2.99    &     0.703  &	26.78  &   $-$1.343  &  $<1.34$  &   $<2.95$  &	 R  &    401  &	       &  rs931701n00  \\
002303.14$+$011533.6     & LBQS 0020+0058        &  0.728  &  18.57  &  $-$24.60  &  2.31  &  30.50  &	  2.92    &     0.754  &	27.02  &   $-$1.338  &  $<0.39$  &   $<0.89$  &	 R  &    426  &	       &  rs931702n00  \\
002311.06$+$003517.4     & LBQS 0020+0018        &  0.422  &  20.24  &  $-$22.08  &  0.47  &  29.33  &	  3.05    &     0.647  &	26.46  &   $-$1.100  &  $<0.50$  &   $<3.27$  &	 R  &    408  &	       &  rs931702n00  \\
003242.74$+$003110.8     &                       &  0.361  &  18.79  &  $-$22.76  &  1.81  &  29.77  &	  4.07    &     0.826  &	26.43  &   $-$1.282  &  $<0.41$  &   $<1.00$  &	 R  &    610  &	       &  rs931702n00  \\
003333.61$-$001858.1     &                       &  0.690  &  19.56  &  $-$23.80  &  1.06  &  30.12  &	  5.21    &     1.31   &	27.21  &   $-$1.116  &  $<0.45$  &   $<1.91$  &	 R  &    626  &	       &  rs931702n00  \\
003340.21$-$005525.6     &                       &  0.937  &  18.23  &  $-$25.56  &  3.28  &  30.88  &	  3.74    &     1.08   &	27.39  &   $-$1.337  &  $<0.45$  &   $<0.79$  &	 R  &    635  &	       &  rs931702n00  \\
003431.74$-$001312.7     &                       &  0.381  &  18.63  &  $-$23.25  &  2.09  &  29.88  &	  2.69    &     0.055  &	25.30  &   $-$1.757  &  $<0.41$  &   $<0.73$  &	 R  &    571  &	       &  rs931702n00  \\
004319.74$+$005115.4     & UM 269                &  0.308  &  18.55  &  $-$22.58  &  2.26  &  29.72  &	  4.62    &     0.901  &	26.32  &   $-$1.305  &     1.60  &      3.22  &	 R  &    320  &	       &  rs931702n00  \\
004613.53$+$010425.7\bq  & UM 275                &  2.152  &  18.36  &  $-$27.29  &  3.31  &  31.56  &  $<0.131$  &   $<0.062$ &     $<26.83$  &  $<-1.816$  &     3.04  &      7.58  &  P  &  10718  &   0.0  &  rp700377n00  \\
004704.46$+$005228.2     &                       &  1.620  &  18.74  &  $-$25.95  &  1.57  &  31.01  &	  0.362   &     0.142  &        26.97  &   $-$1.553  &  $<0.38$  &   $<1.65$  &	 P  &  10718  &  17.2  &  rp700377n00  \\
005121.25$+$004521.6     & LBQS 0048+0029        &  0.727  &  18.91  &  $-$24.52  &  1.67  &  30.36  &  $<0.347$  &   $<0.089$ &     $<26.09$  &  $<-1.640$  &  $<0.43$  &   $<1.07$  &  H  &  12142  &  19.4  &  rh701865n00  \\
005202.40$+$010129.1     & LBQS 0049+0045        &  2.272  &  17.46  &  $-$28.09  &  5.75  &  31.84  &  $<0.436$  &   $<0.213$ &     $<27.41$  &  $<-1.702$  &  $<0.28$  &   $<0.38$  &  H  &  12142  &   0.0  &  rh701865n00  \\
005235.25$+$010227.9     &                       &  1.390  &  18.79  &  $-$25.84  &  1.87  &  30.97  &  $<0.289$  &   $<0.103$ &     $<26.71$  &  $<-1.635$  &  $<0.25$  &   $<0.86$  &  H  &  12142  &   8.2  &  rh701865n00  \\
005441.19$+$000110.6     & LBQS 0052-0015        &  0.647  &  18.60  &  $-$24.44  &  2.14  &  30.37  &	  4.50    &     1.11   &	27.08  &   $-$1.262  &     3.08  &      6.25  &	 R  &    371  &	       &  rs931703n00  \\
010226.32$-$003904.5     & DMS 0059-0055         &  0.294  &  16.78  &  $-$24.33  &  11.5  &  30.39  &	  7.10    &     1.37   &	26.47  &   $-$1.506  &  $<0.46$  &   $<0.17$  &	 R  &    393  &	       &  rs931703n00  \\
010603.86$+$010506.2     &                       &  1.611  &  19.33  &  $-$25.64  &  1.17  &  30.88  &  $<0.647$  &   $<0.252$ &     $<27.21$  &  $<-1.407$  &  $<0.44$  &   $<2.57$  &  H  &  12672  &  11.7  &  rh703871n00  \\
010614.39$+$011409.3     &                       &  1.356  &  19.69  &  $-$24.88  &  0.79  &  30.57  &  $<0.312$  &   $<0.110$ &     $<26.71$  &  $<-1.480$  &  $<0.41$  &   $<3.21$  &  H  &  12672  &  13.9  &  rh703871n00  \\
010648.02$+$004627.8     & LBQS 0104+0030        &  1.883  &  18.59  &  $-$26.49  &  1.91  &  31.22  &  $<0.364$  &   $<0.157$ &     $<27.13$  &  $<-1.568$  &  $<0.45$  &   $<1.71$  &  H  &  12672  &  17.1  &  rh703871n00  \\
011254.91$+$000313.0     & PB 06317              &  0.239  &  18.08  &  $-$22.49  &  3.47  &  29.68  &	  5.89    &     1.09   &	26.18  &   $-$1.345  &  $<0.45$  &   $<0.56$  &	 R  &    419  &	       &  rs931704n00  \\
011757.79$-$010809.9     &                       &  0.707  &  19.27  &  $-$23.62  &  1.18  &  30.19  &	  1.04    &     0.266  &	26.54  &   $-$1.400  &  $<0.48$  &   $<2.57$  &	 H  &  16618  &  15.9  &  rh702758a01  \\
011758.83$+$002021.4     &                       &  0.613  &  19.02  &  $-$23.69  &  1.55  &  30.18  &	  4.47    &     1.07   &	27.02  &   $-$1.212  &  $<0.41$  &   $<1.47$  &	 R  &    406  &	       &  rs931704n00  \\
011817.10$-$011421.9     &                       &  0.668  &  19.94  &  $-$23.27  &  0.63  &  29.86  &  $<0.288$  &   $<0.072$ &     $<25.92$  &  $<-1.513$  &  $<0.47$  &   $<3.02$  &  H  &  16618  &  16.9  &  rh702758a01  \\
011824.29$-$010832.0     &                       &  0.967  &  19.35  &  $-$24.21  &  0.96  &  30.37  &  $<0.296$  &   $<0.087$ &     $<26.33$  &  $<-1.552$  &  $<0.49$  &   $<3.17$  &  H  &  16618  &  11.1  &  rh702758a01  \\
011827.98$-$005239.8     & UM 314                &  2.190  &  18.30  &  $-$27.17  &  2.75  &  31.49  &	  0.319   &     0.152  &	27.24  &   $-$1.634  &  $<0.42$  &   $<1.22$  &	 H  &  16618  &   9.7  &  rh702758a01  \\
011855.75$-$004032.4     &                       &  0.745  &  18.66  &  $-$24.16  &  1.92  &  30.44  &  $<0.292$  &   $<0.076$ &     $<26.04$  &  $<-1.690$  &  $<0.44$  &   $<1.61$  &  H  &  16618  &  19.5  &  rh702758a01  \\
011922.85$-$004419.8     & NGC 0450:[GMS97] 120  &  1.054  &  18.68  &  $-$25.40  &  2.00  &  30.76  &	  1.49    &     0.456  &	27.12  &   $-$1.398  &  $<0.43$  &   $<1.14$  &	 H  &  16618  &  17.5  &  rh702758a01  \\
012602.20$-$001924.1     & NGC 0450:[GMS97] 099  &  1.761  &  18.81  &  $-$26.36  &  1.81  &  31.14  &	  0.213   &     0.088  &	26.83  &   $-$1.657  &  $<2.12$  &   $<7.77$  &	 P  &  13057  &  11.0  &  rp800645n00  \\
013041.65$+$004000.2     &                       &  0.655  &  18.47  &  $-$24.42  &  2.49  &  30.44  &	  3.30    &     0.815  &	26.96  &   $-$1.338  &  $<0.47$  &   $<0.99$  &	 R  &    430  &	       &  rs931705n00  \\
013418.18$+$001536.7     & UM 341                &  0.399  &  17.13  &  $-$24.73  &  8.33  &  30.52  &	  12.2    &     2.54   &	27.01  &   $-$1.349  &  $<0.61$  &   $<0.30$  &	 R  &    386  &	       &  rs931705n00  \\
014017.07$-$005002.9     & UM 357                &  0.335  &  16.70  &  $-$24.67  &  12.4  &  30.54  &	  2.12    &     4.23   &	27.07  &   $-$1.331  &  $<0.45$  &   $<0.16$  &	 R  &    230  &	       &  rs931705n00  \\
014544.09$-$003828.9\bq  &                       &  1.851  &  20.05  &  $-$25.53  &  0.74  &  30.79  &  $<0.338$  &   $<0.144$ &     $<27.08$  &  $<-1.424$  &  $<0.44$  &   $<3.93$  &  P  &   6032  &  13.7  &  rp701219n00  \\
014618.05$-$005131.9     &                       &  0.832  &  19.07  &  $-$24.53  &  1.47  &  30.42  &  $<0.264$  &   $<0.072$ &     $<26.12$  &  $<-1.654$  &  $<0.45$  &   $<1.52$  &  P  &   6032  &  12.4  &  rp701219n00  \\
014619.97$-$004628.8     & UM 368                &  3.175  &  19.02  &  $-$27.10  &  1.32  &  31.45  &	  0.139   &     0.087  &	27.26  &   $-$1.605  &  $<0.43$  &   $<3.07$  &	 P  &   6032  &   7.8  &  rp701219n00  \\
014705.43$-$004148.9\bq  &                       &  2.108  &  20.46  &  $-$26.15  &  1.00  &  31.02  &  $<0.206$  &   $<0.096$ &     $<27.01$  &  $<-1.543$  &  $<0.44$  &   $<2.97$  &  P  &   6032  &   6.9  &  rp701219n00  \\
014942.50$+$001501.7     &                       &  0.552  &  16.81  &  $-$25.68  &  11.4  &  30.95  &	  6.21    &     1.44   &	27.05  &   $-$1.497  &  $<0.41$  &   $<0.18$  &	 R  &    434  &	       &  rs931705n00  \\
015248.42$+$011442.7     &                       &  1.171  &  19.15  &  $-$25.01  &  1.23  &  30.64  &  $<0.431$  &   $<0.140$ &     $<26.70$  &  $<-1.515$  &  $<0.45$  &   $<2.18$  &  H  &  15850  &  14.2  &  rh800731n00  \\
015258.67$+$010507.3     &                       &  0.646  &  19.79  &  $-$23.09  &  0.72  &  29.89  &	  0.427   &     0.105  &	26.06  &   $-$1.473  &  $<0.43$  &   $<2.99$  &	 H  &  15850  &   6.3  &  rh800731n00  \\
015309.13$+$005250.1     &                       &  1.162  &  19.60  &  $-$24.81  &  0.97  &  30.53  &  $<0.314$  &   $<0.101$ &     $<26.55$  &  $<-1.528$  &  $<0.46$  &   $<2.63$  &  H  &  15850  &  10.5  &  rh800731n00  \\
015906.38$+$002343.9     &                       &  2.321  &  19.30  &  $-$26.33  &  1.10  &  31.14  &  $<0.201$  &   $<0.099$ &     $<27.09$  &  $<-1.552$  &  $<0.46$  &   $<3.28$  &  P  &   6155  &  11.0  &  rp700225n00  \\
015910.05$+$010514.5     &                       &  0.217  &  18.04  &  $-$22.56  &  3.61  &  29.61  &	  5.12    &     0.931  &	26.02  &   $-$1.378  &  $<0.44$  &   $<0.41$  &	 R  &    409  &	       &  rs931706n00  \\
015935.48$+$000401.2     &                       &  3.278  &  20.05  &  $-$26.06  &  0.53  &  31.07  &  $<0.267$  &   $<0.170$ &     $<27.58$  &  $<-1.339$  &  $<0.42$  &   $<8.49$  &  P  &   6155  &  19.7  &  rp700225n00  \\
015938.08$+$002639.2     &                       &  1.606  &  18.88  &  $-$25.90  &  1.48  &  30.98  &  $<0.284$  &   $<0.111$ &     $<26.85$  &  $<-1.584$  &  $<0.42$  &   $<1.89$  &  P  &   6155  &   4.5  &  rp700225n00  \\
015945.92$+$000521.4     &                       &  0.389  &  19.18  &  $-$22.61  &  1.26  &  29.68  &	  1.47    &     0.305  &	26.06  &   $-$1.388  &  $<0.42$  &   $<1.39$  &	 P  &   6155  &  18.1  &  rp700225n00  \\
015950.24$+$002340.8     & MRK 1014              &  0.163  &  16.18  &  $-$23.40  &  20.0  &  30.10  &	  9.87    &     1.71   &	26.03  &   $-$1.562  &    24.08  &      5.56  &	 P  &   6155  &   0.0  &  rp700225n00  \\
020047.38$+$003110.1     &                       &  1.620  &  18.45  &  $-$26.21  &  1.97  &  31.11  &	  0.463   &     0.181  &	27.08  &   $-$1.550  &  $<0.45$  &   $<1.57$  &	 P  &   6155  &  16.2  &  rp700225n00  \\
020115.53$+$003135.1     &                       &  0.362  &  18.60  &  $-$22.90  &  2.16  &  29.85  &	  4.25    &     0.863  &	26.45  &   $-$1.304  &  $<0.43$  &   $<0.92$  &	 P  &   1644  &   3.4  &  rp700972n00  \\
021102.72$-$000910.3     &                       &  4.900  &  20.00  &  $-$26.85  &  0.48  &  31.30  &	  0.031   &     0.027  &	27.06  &   $-$1.628  &  $<0.44$  &  $<10.35$  &	 C  &   4888  &   0.0  &  700271       \\
021359.78$+$004226.7     &                       &  0.182  &  18.14  &  $-$21.94  &  3.30  &  29.41  &	  12.0    &     2.12   &	26.22  &   $-$1.226  &     3.90  &      4.39  &	 R  &    235  &	       &  rs931706n00  \\
021425.69$-$003859.7     &                       &  1.726  &  18.06  &  $-$26.73  &  2.88  &  31.33  &  $<0.543$  &   $<0.221$ &     $<27.21$  &  $<-1.580$  &  $<0.46$  &   $<1.14$  &  P  &   3484  &   7.4  &  rp701229n00  \\
021449.10$-$005040.9     &                       &  1.603  &  19.37  &  $-$25.30  &  0.86  &  30.74  &  $<0.501$  &   $<0.195$ &     $<27.10$  &  $<-1.400$  &  $<0.41$  &   $<3.16$  &  P  &   3484  &   5.9  &  rp701229n00  \\
021457.21$-$010323.9     &                       &  1.770  &  19.53  &  $-$25.60  &  0.94  &  30.86  &  $<0.503$  &   $<0.208$ &     $<27.21$  &  $<-1.403$  &  $<0.45$  &   $<3.35$  &  P  &   3484  &  18.2  &  rp701229n00  \\
023325.33$+$002914.8     &                       &  2.014  &  18.45  &  $-$26.76  &  2.29  &  31.35  &	  0.381   &     0.172  &	27.22  &   $-$1.583  &  $<0.43$  &   $<1.48$  &	 P  &  28216  &  16.1  &  rp800482n00  \\
023333.23$+$010333.0     &                       &  2.060  &  18.58  &  $-$26.80  &  2.09  &  31.33  &  $<0.108$  &   $<0.049$ &     $<26.70$  &  $<-1.777$  &  $<0.45$  &   $<1.59$  &  P  &  28216  &  18.6  &  rp800482n00  \\
023359.71$+$004938.5     &                       &  2.523  &  17.78  &  $-$27.88  &  4.37  &  31.80  &	  0.286   &     0.151  &	27.34  &   $-$1.713  &  $<0.45$  &   $<0.92$  &	 P  &  28216  &   6.6  &  rp800482n00  \\
024141.52$+$000416.5     & [CLA95] 023907.70-00  &  0.648  &  18.83  &  $-$23.85  &  1.63  &  30.25  &	  0.470   &     0.116  &	26.10  &   $-$1.593  &  $<1.19$  &   $<4.14$  &	 P  &   5471  &  15.5  &  rp150021a02  \\
024240.31$+$005727.1     & E 0240+007            &  0.569  &  16.74  &  $-$25.82  &  12.2  &  31.01  &	  0.164   &     3.85   &	27.51  &   $-$1.343  &     3.58  &      1.50  &	 R  &    234  &	       &  rs931708n00  \\
024304.68$+$000005.4\bq  & [CLA95] 024030.91-00  &  2.003  &  18.53  &  $-$26.85  &  2.51  &  31.38  &	  0.220   &     0.099  &	26.98  &   $-$1.691  &  $<1.25$  &   $<3.95$  &	 P  &   5471  &   6.0  &  rp150021a02  \\
024441.42$+$011434.4     &                       &  0.546  &  20.09  &  $-$23.03  &  0.75  &  29.76  &	  1.34    &     0.308  &	26.37  &   $-$1.299  &  $<0.42$  &   $<2.10$  &	 H  &  27741  &  17.4  &  rh600999n00  \\
025331.93$+$001624.7\bq  & [LCB92] 0250+0004     &  1.825  &  19.01  &  $-$26.03  &  1.34  &  31.04  &  $<0.606$  &   $<0.255$ &     $<27.32$  &  $<-1.428$  &  $<0.44$  &   $<2.40$  &  P  &   2949  &  19.8  &  rp190332n00  \\
025356.07$+$001057.3     & [LCB92] 0251-0001B    &  1.699  &  19.29  &  $-$25.55  &  0.99  &  30.85  &  $<0.607$  &   $<0.245$ &     $<27.24$  &  $<-1.385$  &  $<0.44$  &   $<3.08$  &  P  &   2949  &  13.0  &  rp190332n00  \\
025438.36$+$002132.6     &                       &  2.465  &  19.73  &  $-$26.14  &  0.83  &  31.06  &	  0.081   &     0.042  &	26.76  &   $-$1.650  &  $<0.44$  &   $<4.25$  &	 P  &  11855  &  13.4  &  rp701403n00  \\
025505.66$+$002523.0     & US 3333               &  0.354  &  17.94  &  $-$23.38  &  3.96  &  30.09  &	  4.69    &     0.948  &	26.47  &   $-$1.390  &  $<0.45$  &   $<0.60$  &	 P  &  11855  &  14.7  &  rp701403n00  \\
025513.02$+$000639.4     & [LCB92] 0252-0005     &  1.886  &  19.12  &  $-$25.84  &  1.05  &  30.96  &	  0.309   &     0.133  &	27.06  &   $-$1.495  &  $<0.42$  &   $<2.92$  &	 P  &  11855  &   4.2  &  rp701403n00  \\
025518.58$+$004847.4     &                       &  3.990  &  18.76  &  $-$27.80  &  1.91  &  31.77  &  $<1.46$   &   $<1.080$  &    $<28.52$  &  $<-1.246$  &  $<0.45$  &   $<2.80$  &  P  &    850  &  14.1  &  rp190035n00  \\
025559.92$+$005311.4     & [HB89] 0253+006       &  0.846  &  18.71  &  $-$24.72  &  1.97  &  30.57  &  $<1.56$   &   $<0.429$ &     $<26.90$  &  $<-1.406$  &  $<0.46$  &   $<1.35$  &  P  &    850  &  11.4  &  rp190035n00  \\
025607.24$+$011038.6     & [LCB92] 0253+0058     &  1.349  &  19.19  &  $-$25.59  &  1.45  &  30.83  &  $<1.76$   &   $<0.618$ &     $<27.46$  &  $<-1.294$  &  $<0.45$  &   $<1.78$  &  P  &    850  &  11.9  &  rp190035n00  \\
025702.08$-$010217.2     & [LCB92] 0254-0114     &  0.874  &  19.32  &  $-$24.14  &  1.01  &  30.31  &	  0.340   &     0.095  &	26.28  &   $-$1.546  &  $<0.45$  &   $<2.42$  &	 P  &   2996  &  11.6  &  rp190338n00  \\
025705.87$-$001053.5     & [LCB92] 0254-0022     &  1.586  &  19.43  &  $-$25.48  &  1.03  &  30.81  &  $<1.29$   &   $<0.498$ &     $<27.50$  &  $<-1.272$  &  $<0.44$  &   $<2.86$  &  P  &    949  &  11.4  &  rp190037n00  \\
025713.07$-$010157.8     & [CLB91] 025440.2-011  &  1.869  &  19.31  &  $-$25.82  &  1.02  &  30.94  &	  0.753   &     0.322  &	27.44  &   $-$1.343  &  $<0.45$  &   $<3.13$  &	 P  &   2996  &  13.6  &  rp190338n00  \\
025717.35$-$011413.0     &                       &  1.924  &  19.36  &  $-$25.78  &  0.95  &  30.93  &  $<0.698$  &   $<0.305$ &     $<27.44$  &  $<-1.342$  &  $<0.43$  &   $<3.27$  &  P  &   1996  &  10.6  &  rp190338n00  \\
025735.33$-$001631.4     &                       &  0.362  &  20.45  &  $-$21.91  &  0.39  &  29.11  &  $<1.30$   &   $<0.264$ &     $<25.94$  &  $<-1.218$  &  $<0.45$  &   $<2.43$  &  P  &    949  &   6.2  &  rp190037n00  \\
025743.73$+$011144.4     &                       &  1.705  &  19.25  &  $-$25.84  &  1.34  &  30.99  &  $<1.21$   &   $<0.490$ &     $<27.55$  &  $<-1.320$  &  $<0.45$  &   $<2.47$  &  P  &    986  &  17.8  &  rp190328n00  \\
025751.55$+$002045.4     & [HB89] 0255+001       &  1.500  &  19.81  &  $-$25.24  &  0.88  &  30.70  &  $<0.401$  &   $<0.150$ &     $<26.93$  &  $<-1.447$  &  $<0.44$  &   $<3.14$  &  P  &   2894  &   7.3  &  rp190330n00  \\
025815.54$-$000334.2     & US 3437               &  1.319  &  19.26  &  $-$25.21  &  1.12  &  30.70  &  $<0.488$  &   $<0.169$ &     $<26.88$  &  $<-1.467$  &  $<0.44$  &   $<2.33$  &  P  &   4984  &  19.4  &  rp700393n00  \\
025905.64$+$001121.9     & LBQS 0256-0000        &  3.366  &  17.63  &  $-$28.61  &  5.01  &  32.07  &	  0.960   &     0.625  &	28.16  &   $-$1.499  &    2.59   &      5.29  &	 P  &   4984  &   0.0  &  rp700393n00  \\
025922.64$+$005829.4     & US 3461               &  1.861  &  18.63  &  $-$26.47  &  1.95  &  31.22  &  $<1.19$   &   $<0.508$ &     $<27.63$  &  $<-1.375$  &  $<0.45$  &   $<1.68$  &  P  &    986  &  19.8  &  rp190328n00  \\
030027.11$-$004848.8     & US 3499               &  2.015  &  18.86  &  $-$26.41  &  1.58  &  31.19  &  $<0.958$  &   $<0.431$ &     $<27.62$  &  $<-1.369$  &  $<0.43$  &   $<2.06$  &  P  &   1801  &  10.6  &  rp190340n00  \\
030035.97$-$001533.0     & [LCB92] 0258-0027     &  1.446  &  18.69  &  $-$25.76  &  1.70  &  30.96  &  $<0.826$  &   $<0.301$ &     $<27.20$  &  $<-1.440$  &  $<0.71$  &   $<2.86$  &  P  &   1814  &  10.8  &  rp190334n00  \\
030100.23$+$000429.2     &                       &  0.486  &  19.64  &  $-$22.71  &  0.85  &  29.71  &  $<0.709$  &   $<0.157$ &     $<25.98$  &  $<-1.434$  &  $<0.44$  &   $<2.25$  &  P  &   1814  &  10.6  &  rp190334n00  \\
030335.76$+$004144.8     &                       &  0.669  &  20.24  &  $-$23.57  &  0.58  &  29.83  &  $<0.340$  &   $<0.085$ &     $<25.99$  &  $<-1.472$  &  $<0.45$  &   $<2.19$  &  P  &   5584  &  18.4  &  rp700395n00  \\
030342.79$+$002700.5     & LBQS 0301+0015        &  1.651  &  18.00  &  $-$26.92  &  3.66  &  31.40  &	  0.466   &     0.184  &	27.10  &   $-$1.650  &  $<0.44$  &   $<0.83$  &	 P  &   5584  &   3.8  &  rp700395n00  \\
030422.39$+$002231.7     & LBQS 0301+0010        &  0.638  &  17.27  &  $-$25.57  &  7.31  &  30.89  &	  0.173   &     0.042  &	25.65  &   $-$2.011  &  $<0.43$  &   $<0.30$  &	 P  &   5584  &  10.3  &  rp700395n00  \\
103027.10$+$052455.0     &                       &  6.280  &  19.66  &  $-$27.59  &  0.65  &  31.60  &	  0.020   &     0.021  &	27.11  &   $-$1.723  &  $<0.46$  &   $<9.32$  &	 C  &   7956  &   0.0  &  700605       \\
103457.28$-$010209.0     &                       &  0.328  &  18.37  &  $-$23.20  &  2.66  &  29.85  &	  4.04    &     0.801  &	26.33  &   $-$1.352  &  $<0.43$  &   $<0.56$  &	 R  &    385  &	       &  rs931729n00  \\
103709.80$+$000235.1     &                       &  2.681  &  20.30  &  $-$25.75  &  0.53  &  30.92  &  $<0.145$  &   $<0.080$ &     $<27.11$  &  $<-1.466$  &  $<0.46$  &   $<7.85$  &  P  &  18097  &  11.1  &  rp201243n00  \\
103747.41$-$001643.9     &                       &  1.498  &  20.00  &  $-$25.21  &  0.85  &  30.68  &  $<0.183$  &   $<0.068$ &     $<26.59$  &  $<-1.572$  &  $<0.42$  &   $<3.07$  &  P  &  18097  &  13.5  &  rp201243n00  \\
104152.61$-$001102.1\bq  & 2QZ J104152.5-001102  &  1.703  &  19.45  &  $-$25.58  &  0.94  &  30.83  &  $<0.203$  &   $<0.082$ &     $<26.77$  &  $<-1.559$  &  $<0.45$  &   $<3.14$  &  P  &   7202  &  11.2  &  rp800194n00  \\
104159.13$-$001126.3     & 2QZ J104159.0-001127  &  0.794  &  19.76  &  $-$23.93  &  0.75  &  30.09  &	  0.516   &     0.138  &	26.36  &   $-$1.433  &  $<0.45$  &   $<2.37$  &	 P  &   7202  &   9.6  &  rp800194n00  \\
104243.13$-$001706.0     &                       &  1.972  &  18.27  &  $-$27.01  &  2.85  &  31.43  &	  0.225   &     0.100  &	26.97  &   $-$1.711  &  $<0.46$  &   $<1.20$  &	 P  &   7202  &   2.7  &  rp800194n00  \\
104253.44$-$001300.8     &                       &  2.957  &  18.89  &  $-$27.18  &  1.58  &  31.48  &	  0.296   &     0.174  &	27.52  &   $-$1.520  &  $<0.45$  &   $<2.55$  &	 P  &   7202  &   5.5  &  rp800194n00  \\
104433.04$-$012502.2\bq  &                       &  5.745  &  19.21  &  $-$27.90  &  0.98  &  31.72  &	  0.012   &     0.012  &	26.81  &   $-$1.886  &  $<0.45$  &   $<5.72$  &	 X  &  32613  &   0.0  &  0125300101   \\
104552.42$+$002410.2     &                       &  0.556  &  18.96  &  $-$23.53  &  1.60  &  30.11  &	  3.56    &     0.827  &	26.82  &   $-$1.262  &  $<0.47$  &   $<1.52$  &	 R  &    405  &	       &  rs931729n00  \\
105336.71$-$001727.3     & 2QZ J105336.6-001727  &  0.482  &  19.55  &  $-$22.56  &  0.88  &  29.72  &	  0.750   &     0.166  &	25.99  &   $-$1.429  &  $<0.45$  &   $<2.59$  &	 P  &   4732  &   4.0  &  rp700381n00  \\
105337.95$+$005958.8     &                       &  0.476  &  19.64  &  $-$22.69  &  0.85  &  29.69  &	  3.42    &     0.754  &	26.64  &   $-$1.172  &  $<0.44$  &   $<2.19$  &	 R  &    416  &	       &  rs931730n00  \\
105338.37$-$002804.4     & 2QZ J105338.3-002805  &  1.840  &  18.81  &  $-$26.28  &  1.65  &  31.14  &  $<0.478$  &   $<0.203$ &     $<27.23$  &  $<-1.501$  &  $<0.42$  &   $<1.85$  &  P  &   4732  &  12.0  &  rp700381n00  \\
105414.10$-$001803.6     & 2QZ J105414.1-001804  &  1.163  &  19.39  &  $-$24.87  &  1.05  &  30.57  &  $<0.309$  &   $<0.100$ &     $<26.54$  &  $<-1.544$  &  $<0.44$  &   $<2.39$  &  P  &   4732  &  13.2  &  rp700381n00  \\
110057.71$-$005304.5     &                       &  0.378  &  17.80  &  $-$23.75  &  4.49  &  30.21  &	  6.31    &     1.30   &	26.67  &   $-$1.359  &  $<0.52$  &   $<0.57$  &	 R  &    433  &	       &  rs931730n00  \\
112405.48$+$002401.9     &                       &  0.789  &  19.35  &  $-$24.48  &  1.17  &  30.28  &	  3.13    &     0.836  &	27.13  &   $-$1.208  &  $<0.42$  &   $<1.30$  &	 R  &    421  &	       &  rs931731n00  \\
112956.10$-$014212.4\bq  &                       &  4.850  &  19.20  &  $-$27.63  &  0.97  &  31.62  &  $<0.020$  &   $<0.017$ &     $<26.85$  &  $<-1.829$  &  $<0.41$  &   $<4.65$  &  C  &   4818  &   0.0  &  700272       \\
113541.20$+$002235.2     & [VCV96] Q 1133+0039   &  0.175  &  18.11  &  $-$21.81  &  3.39  &  29.39  &	  4.94    &     0.865  &	25.80  &   $-$1.379  &  $<0.42$  &   $<0.49$  &	 R  &    427  &	       &  rs931731n00  \\
114612.09$+$002105.1     & 2QZ J114612.0+002104  &  1.229  &  19.19  &  $-$25.03  &  1.21  &  30.67  &	  0.343   &     0.114  &	26.65  &   $-$1.545  &  $<0.49$  &   $<2.63$  &	 P  &   5837  &  10.2  &  rp201242n00  \\
114647.72$+$001350.6     & LBQS 1144+0030        &  0.940  &  18.22  &  $-$25.47  &  2.95  &  30.83  &	  0.719   &     0.208  &	26.68  &   $-$1.594  &  $<0.43$  &   $<0.82$  &	 P  &   5837  &   3.1  &  rp201242n00  \\
114733.67$+$000811.5     & 2QZ J114733.6+000811  &  1.077  &  19.55  &  $-$24.54  &  0.89  &  30.43  &	  0.586   &     0.182  &	26.74  &   $-$1.416  &  $<0.46$  &   $<2.82$  &	 P  &   5837  &  15.1  &  rp201242n00  \\
115758.73$-$002220.8     &                       &  0.260  &  17.38  &  $-$23.46  &  6.61  &  30.04  &	  2.64    &     0.497  &	25.91  &   $-$1.583  &  $<0.46$  &   $<0.28$  &	 R  &    359  &	       &  rs931732n00  \\
120014.08$-$004638.7     &                       &  0.179  &  18.48  &  $-$21.66  &  2.41  &  29.26  &	  11.1    &     1.96   &	26.17  &   $-$1.186  &  $<0.44$  &   $<0.61$  &	 R  &    366  &	       &  rs931733n00  \\
120441.73$-$002149.6     &                       &  5.030  &  19.10  &  $-$27.79  &  1.09  &  31.68  &	  0.038   &     0.034  &	27.17  &   $-$1.730  &  $<0.45$  &   $<4.69$  &	 C  &   6272  &   0.0  &  700373       \\
120823.82$+$001027.7     &                       &  5.273  &  20.50  &  $-$26.47  &  0.30  &  31.15  &	  0.011   &     0.010  &	26.69  &   $-$1.713  &  $<0.46$  &  $<18.05$  &	 C  &   4619  &   0.0  &  700273       \\
123514.95$+$004740.7     &                       &  1.876  &  19.92  &  $-$25.70  &  0.86  &  30.87  &  $<0.180$  &   $<0.077$ &     $<26.82$  &  $<-1.554$  &  $<0.48$  &   $<3.71$  &  P  &  23511  &  18.6  &  rp700355n00  \\
123552.74$+$005029.6     &                       &  2.241  &  19.29  &  $-$26.26  &  1.10  &  31.11  &	  0.078   &     0.378  &	27.65  &   $-$1.329  &  $<0.42$  &   $<2.95$  &	 P  &  23511  &   8.8  &  rp700355n00  \\
123659.89$+$004212.1     & 2QZ J123659.9+004211  &  2.378  &  19.06  &  $-$26.47  &  1.32  &  31.24  &	  0.177   &     0.089  &	27.07  &   $-$1.601  &  $<0.44$  &   $<2.94$  &	 P  &  23511  &  15.3  &  rp700355n00  \\
123724.52$+$010615.5     & LBQS 1234+0122        &  2.023  &  18.23  &  $-$27.17  &  3.02  &  31.47  &  $<0.303$  &   $<0.049$ &     $<26.68$  &  $<-1.840$  &  $<0.42$  &   $<1.01$  &  P  &  23511  &  19.1  &  rp700355n00  \\
123729.74$+$005433.6     &                       &  3.279  &  18.99  &  $-$27.26  &  1.45  &  31.51  &	  0.201   &     0.109  &	27.39  &   $-$1.583  &  $<0.46$  &   $<3.03$  &	 P  &  23511  &  16.4  &  rp700355n00  \\
124540.99$-$002744.9     & LBQS 1243-0011        &  1.687  &  18.98  &  $-$26.23  &  1.72  &  31.08  &	  1.01    &     0.407  &	27.46  &   $-$1.392  &  $<0.48$  &   $<1.80$  &	 P  &  16761  &   7.8  &  rp600532n00  \\
124555.12$-$003735.5     & 2QZ J124555.1-003736  &  1.042  &  18.94  &  $-$24.92  &  1.50  &  30.63  &	  0.482   &     0.147  &	26.62  &   $-$1.539  &  $<0.45$  &   $<1.82$  &	 P  &  16761  &  15.2  &  rp600532n00  \\
130018.84$+$000756.5     &                       &  1.305  &  20.36  &  $-$25.18  &  0.82  &  30.56  &  $<0.309$  &   $<0.106$ &     $<26.67$  &  $<-1.493$  &  $<0.43$  &   $<2.30$  &  H  &   7247  &   0.9  &  rh201701n00  \\
130608.26$+$035626.3     &                       &  5.990  &  19.55  &  $-$27.62  &  0.72  &  31.61  &	  0.044   &     0.046  &	27.42  &   $-$1.611  &  $<0.41$  &   $<7.24$  &	 C  &   8158  &   0.0  &  700606       \\
130713.25$-$003601.7     &                       &  0.170  &  17.96  &  $-$21.78  &  3.89  &  29.43  &	  7.78    &     1.36   &	25.97  &   $-$1.327  &  $<0.41$  &   $<0.46$  &	 R  &    251  &	       &  rs931735n00  \\
131131.03$-$010332.3     &                       &  1.304  &  18.94  &  $-$25.49  &  1.50  &  30.82  &  $<0.150$  &   $<0.052$ &     $<26.35$  &  $<-1.714$  &  $<0.48$  &   $<1.93$  &  P  &  13949  &  16.9  &  rp800248n00  \\
131208.68$-$010710.1     &                       &  0.805  &  19.61  &  $-$23.78  &  0.87  &  30.17  &	  0.602   &     0.162  &	26.44  &   $-$1.432  &  $<0.47$  &   $<2.89$  &	 P  &  13949  &  16.1  &  rp800248n00  \\
132020.27$-$003428.2     & LBQS 1317-0018        &  0.353  &  18.94  &  $-$22.52  &  1.58  &  29.69  &	  3.86    &     0.780  &	26.39  &   $-$1.269  &  $<0.70$  &   $<2.04$  &	 R  &    273  &	       &  rs931736n00  \\
134044.52$-$004516.6     & LBQS 1338-0030        &  0.384  &  17.63  &  $-$23.97  &  5.25  &  30.29  &	  2.99    &     0.618  &	26.36  &   $-$1.509  &  $<0.43$  &   $<0.40$  &	 R  &    310  &	       &  rs931737n00  \\
134113.93$-$005315.0     & LBQS 1338-0038        &  0.237  &  18.15  &  $-$22.42  &  3.25  &  29.65  &	  6.06    &     1.12   &	26.18  &   $-$1.330  &     5.38  &      6.96  &	 R  &    343  &	       &  rs931737n00  \\
134233.70$-$001148.1     & 2QZ J134233.7-001148  &  0.516  &  19.67  &  $-$22.73  &  0.85  &  29.76  &	  0.213   &     0.048  &	25.52  &   $-$1.630  &  $<0.47$  &   $<2.69$  &	 P  &  27254  &  17.6  &  rp701000a01  \\
134256.51$+$000057.2     & F864:072              &  0.804  &  19.48  &  $-$23.84  &  1.00  &  30.23  &	  1.29    &     0.347  &	26.77  &   $-$1.328  &  $<0.47$  &   $<2.74$  &	 P  &  27254  &  19.5  &  rp701000a01  \\
134325.84$-$001612.2     & 2QZ J134325.8-001612  &  1.514  &  19.45  &  $-$25.40  &  1.01  &  30.77  &  $<0.084$  &   $<0.032$ &     $<26.26$  &  $<-1.729$  &  $<0.42$  &   $<2.64$  &  P  &  27254  &   4.5  &  rp701000a01  \\
134425.94$-$000056.2     &                       &  1.097  &  18.99  &  $-$24.92  &  1.37  &  30.63  &	  0.253   &     0.079  &	26.39  &   $-$1.627  &  $<0.44$  &   $<2.01$  &	 P  &  27254  &  17.7  &  rp701000a01  \\
134459.44$-$001559.5     & LBQS 1342-0000        &  0.245  &  18.21  &  $-$22.48  &  3.08  &  29.65  &	  2.06    &     0.383  &	25.75  &   $-$1.499  &  $<0.45$  &   $<0.60$  &	 P  &  21103  &  14.6  &  rp800369a01  \\
134507.93$-$001900.9     &                       &  0.419  &  20.61  &  $-$22.48  &  0.34  &  29.18  &	  0.370   &     0.078  &	25.54  &   $-$1.395  &  $<0.44$  &   $<1.95$  &	 P  &   5048  &  17.8  &  rp800369n00  \\
135128.32$+$010338.5     & LBQS 1348+0118        &  1.085  &  17.65  &  $-$26.51  &  5.57  &  31.23  &	  3.82    &     1.19   &	27.56  &   $-$1.409  &  $<0.47$  &   $<0.48$  &	 R  &    377  &	       &  rs931737n00  \\
141637.44$+$003352.2     & 2QZ J141637.4+003351  &  0.434  &  19.77  &  $-$22.60  &  0.74  &  29.55  &	  3.64    &     0.779  &	26.57  &   $-$1.143  &  $<0.41$  &   $<1.78$  &	 R  &    404  &	       &  rs931739n00  \\
142441.21$-$000727.2     &                       &  0.318  &  19.13  &  $-$22.50  &  1.32  &  29.52  &	  8.36    &     1.64   &	26.61  &   $-$1.115  &  $<0.45$  &   $<1.06$  &	 R  &    336  &	       &  rs931739n00  \\
142658.57$-$002056.4     & EQS B1424-0007        &  0.627  &  17.24  &  $-$25.52  &  7.91  &  30.91  &	  4.10    &     0.995  &	27.01  &   $-$1.497  &  $<0.42$  &   $<0.29$  &	 R  &    328  &	       &  rs931739n00  \\
143143.80$-$005011.4     & LBQS 1429-0036        &  1.190  &  18.59  &  $-$25.84  &  2.46  &  30.96  &  $<0.146$  &   $<0.048$ &     $<26.24$  &  $<-1.809$  &  $<0.55$  &   $<1.29$  &  H  &  33627  &  19.4  &  rh800679n00  \\
152241.31$-$004214.9     &                       &  1.081  &  19.59  &  $-$24.84  &  1.05  &  30.51  &  $<2.11$   &   $<0.655$ &     $<27.30$  &  $<-1.231$  &  $<0.41$  &   $<1.93$  &  H  &   1434  &  16.9  &  rh701907n00  \\
152316.31$-$004150.5     &                       &  0.444  &  18.62  &  $-$23.28  &  2.15  &  30.03  &  $<1.57$   &   $<0.337$ &     $<26.23$  &  $<-1.460$  &  $<0.45$  &   $<1.09$  &  H  &   1434  &   8.5  &  rh701907n00  \\
152348.99$-$004701.8\bq  &                       &  3.293  &  18.61  &  $-$27.77  &  2.40  &  31.73  &  $<2.10$   &   $<1.34$  &     $<28.48$  &  $<-1.248$  &  $<0.41$  &   $<1.74$  &  H  &   1434  &   2.0  &  rh701907n00  \\
153259.96$-$003944.1     &                       &  4.620  &  19.40  &  $-$27.35  &  0.83  &  31.50  &  $<0.020$  &   $<0.017$ &     $<26.82$  &  $<-1.797$  &  $<0.45$  &   $<5.92$  &  C  &   5087  &   0.0  &  700275       \\
160501.21$-$011220.0\bq  &                       &  4.920  &  19.40  &  $-$27.46  &  0.83  &  31.55  &  $<0.020$  &   $<0.018$  &    $<26.87$  &  $<-1.793$  &  $<0.45$  &   $<6.17$  &  C  &   4641  &   0.0  &  700276       \\
165338.68$+$634010.5     &                       &  0.279  &  18.51  &  $-$22.55  &  2.34  &  29.65  &	  0.826   &     0.158  &	25.48  &   $-$1.601  &  $<1.50$  &   $<2.48$  &	 R  &   2839  &	       &  rs930624n00  \\
165537.78$+$624739.0     &                       &  0.597  &  19.50  &  $-$23.49  &  0.98  &  29.96  &	  0.453   &     0.108  &	26.00  &   $-$1.519  &  $<1.50$  &   $<6.00$  &	 P  &  17735  &  18.1  &  rp900169n00  \\
165627.31$+$623226.8     &                       &  0.185  &  18.73  &  $-$21.63  &  1.91  &  29.19  &	  0.186   &     0.033  &	24.43  &   $-$1.829  &  $<1.50$  &   $<2.33$  &	 P  &  17735  &   5.3  &  rp900169n00  \\
165951.98$+$622758.6     &                       &  1.325  &  19.41  &  $-$24.98  &  0.91  &  30.61  &  $<0.181$  &   $<0.063$ &     $<26.45$  &  $<-1.598$  &  $<1.50$  &  $<10.10$  &  P  &  17735  &  19.1  &  rp900169n00  \\
165958.93$+$620218.1     &                       &  0.232  &  18.37  &  $-$22.19  &  2.65  &  29.54  &	  1.41    &     0.259  &	25.53  &   $-$1.539  &  $<1.50$  &   $<2.29$  &	 R  &   1975  &	       &  rs930624n00  \\
170049.55$+$643159.3     &                       &  1.258  &  18.84  &  $-$25.23  &  1.39  &  30.75  &	  0.362   &     0.122  &	26.70  &   $-$1.557  &  $<1.50$  &   $<7.10$  &	 P  &  16073  &  19.9  &  rp701121n00  \\
170113.94$+$635243.0     &                       &  1.762  &  18.08  &  $-$26.76  &  2.81  &  31.33  &	  0.211   &     0.087  &	26.82  &   $-$1.731  &  $<1.50$  &   $<3.80$  &	 P  &  27387  &  19.3  &  rp701457n00  \\
170406.14$+$604753.7     & CRSS J1704.1+6047     &  1.364  &  18.68  &  $-$25.80  &  1.71  &  30.91  &	  0.724   &     0.255  &	27.09  &   $-$1.469  &  $<1.50$  &   $<5.05$  &	 P  &  15372  &   5.6  &  rp701439n00  \\
170438.97$+$585748.7     &                       &  0.918  &  19.25  &  $-$24.75  &  1.47  &  30.51  &	  1.61    &     0.462  &	27.01  &   $-$1.344  &  $<1.50$  &   $<5.22$  &	 R  &   1160  &	       &  rs930728n00  \\
170441.37$+$604430.4     & [HB89] 1704+608       &  0.372  &  16.13  &  $-$25.77  &  21.0  &  30.86  &	  4.30    &     0.881  &	26.49  &   $-$1.680  &  $<1.50$  &   $<0.24$  &	 P  &  15372  &   0.0  &  rp701439n00  \\
170611.40$+$610052.9     &                       &  2.061  &  19.36  &  $-$26.05  &  1.10  &  31.05  &  $<0.249$  &   $<0.114$ &     $<27.06$  &  $<-1.530$  &  $<1.50$  &  $<10.53$  &  P  &  15372  &  19.8  &  rp701439n00  \\
170817.87$+$615448.6     &                       &  1.419  &  18.53  &  $-$26.38  &  2.54  &  31.11  &	  1.26    &     0.456  &	27.37  &   $-$1.438  &  $<1.50$  &   $<3.25$  &	 R  &   1808  &	       &  rs930624n00  \\
170839.83$+$652137.7     &                       &  1.397  &  18.60  &  $-$25.81  &  1.91  &  30.98  &	  0.636   &     0.228  &	27.05  &   $-$1.506  &  $<1.50$  &   $<5.29$  &	 R  &   3171  &	       &  rs930521n00  \\
170918.26$+$622516.1     &                       &  1.406  &  19.15  &  $-$25.57  &  1.35  &  30.83  &	  0.684   &     0.245  &	27.09  &   $-$1.436  &  $<1.50$  &   $<6.73$  &	 R  &   2079  &	       &  rs930624n00  \\
170930.80$+$641002.9     &                       &  1.592  &  18.64  &  $-$26.24  &  1.97  &  31.10  &	  0.721   &     0.277  &	27.25  &   $-$1.479  &  $<1.50$  &   $<4.86$  &	 R  &   3337  &	       &  rs930624n00  \\
170937.63$+$640615.0     &                       &  1.119  &  18.85  &  $-$25.01  &  1.32  &  30.63  &	  0.156   &     0.049  &	26.21  &   $-$1.699  &  $<1.50$  &   $<6.55$  &	 P  &  14534  &  18.5  &  rp800512n00  \\
170956.01$+$573225.4     &                       &  0.522  &  19.06  &  $-$23.58  &  1.51  &  30.02  &	  2.14    &     0.487  &	26.53  &   $-$1.340  &  $<0.43$  &   $<1.15$  &	 R  &   1612  &	       &  rs930728n00  \\
171007.50$+$641240.9     &                       &  1.376  &  18.25  &  $-$26.23  &  2.60  &  31.10  &	  0.767   &     0.272  &	27.12  &   $-$1.528  &  $<1.50$  &   $<3.47$  &	 P  &  14534  &  17.9  &  rp800512n00  \\
171009.85$+$642256.7     &                       &  0.256  &  19.93  &  $-$21.70  &  0.63  &  29.00  &	  0.599   &     0.112  &	25.25  &   $-$1.439  &  $<1.50$  &   $<4.48$  &	 R  &   3498  &	       &  rs930624n00  \\
171015.72$+$634806.1     &                       &  1.786  &  19.32  &  $-$25.70  &  1.01  &  30.90  &	  0.659   &     0.274  &	27.33  &   $-$1.369  &  $<1.50$  &  $<10.41$  &	 R  &   3194  &	       &  rs930624n00  \\
171022.61$+$640602.9     &                       &  1.248  &  18.89  &  $-$25.55  &  1.72  &  30.84  &	  0.162   &     0.054  &	26.34  &   $-$1.727  &  $<1.50$  &   $<5.19$  &	 P  &  14534  &  13.8  &  rp800512n00  \\
171117.66$+$584123.7     &                       &  0.616  &  19.46  &  $-$23.67  &  1.09  &  30.03  &	  0.731   &     0.176  &	26.24  &   $-$1.456  &  $<0.43$  &   $<1.57$  &	 P  &   7902  &  11.1  &  rp900171n00  \\
171126.94$+$585544.2     &                       &  0.537  &  19.54  &  $-$22.97  &  0.93  &  29.84  &	  0.592   &     0.136  &	26.00  &   $-$1.473  &  $<1.50$  &   $<7.57$  &	 P  &   7902  &   7.7  &  rp900171n00  \\
171127.49$+$641722.4     &                       &  0.798  &  19.37  &  $-$23.84  &  1.07  &  30.25  &	  0.710   &     0.191  &	26.50  &   $-$1.440  &  $<1.50$  &   $<8.63$  &	 P  &  14534  &  15.3  &  rp800512n00  \\
171144.82$+$584919.6     &                       &  1.532  &  19.29  &  $-$25.39  &  0.99  &  30.77  &  $<0.133$  &   $<0.050$ &     $<26.47$  &  $<-1.648$  &  $<1.50$  &   $<9.71$  &  P  &   7902  &   3.4  &  rp900171n00  \\
171207.44$+$584754.5     &                       &  0.269  &  18.55  &  $-$22.32  &  2.25  &  29.60  &	  2.88    &     0.546  &	25.99  &   $-$1.388  &  $<0.43$  &   $<0.80$  &	 P  &   7902  &   2.5  &  rp900171n00  \\
171300.68$+$572530.2     &                       &  0.360  &  18.85  &  $-$22.92  &  1.71  &  29.74  &	  0.553   &     0.112  &	25.56  &   $-$1.606  &     1.25  &      2.63  &	 P  &   5311  &  18.6  &  rp600425n00  \\
171330.21$+$644253.0     &                       &  1.048  &  18.11  &  $-$25.69  &  2.95  &  30.93  &	  0.725   &     0.222  &	26.80  &   $-$1.583  &  $<1.50$  &   $<3.00$  &	 R  &   3759  &	       &  rs930521n00  \\
171352.42$+$584201.2     &                       &  0.521  &  18.77  &  $-$24.08  &  2.06  &  30.16  &  $<0.378$  &   $<0.086$ &     $<25.78$  &  $<-1.682$  &  $<1.50$  &   $<2.54$  &  P  &   7902  &  15.7  &  rp900171n00  \\
171359.52$+$640939.5     &                       &  1.363  &  19.02  &  $-$25.61  &  1.46  &  30.84  &	  0.286   &     0.101  &	26.68  &   $-$1.598  &  $<1.50$  &   $<6.03$  &	 P  &  14534  &  11.6  &  rp800512n00  \\
171441.87$+$644155.2     & HS 1714+6445          &  0.285  &  18.88  &  $-$22.01  &  1.66  &  29.52  &	  6.13    &     1.17   &	26.37  &   $-$1.210  &  $<1.50$  &   $<4.27$  &	 R  &   3759  &	       &  rs930521n00  \\
171522.17$+$573626.6     &                       &  2.030  &  17.53  &  $-$27.97  &  6.31  &  31.80  &	  0.871   &     0.394  &	27.59  &   $-$1.614  &  $<0.44$  &   $<0.51$  &	 P  &   5311  &  10.6  &  rp600425n00  \\
171622.57$+$551213.4     &                       &  0.323  &  19.04  &  $-$22.43  &  1.44  &  29.57  &	  2.99    &     0.591  &	26.18  &   $-$1.300  &  $<0.45$  &   $<1.16$  &	 R  &   1233  &	       &  rs930729n00  \\
171750.59$+$581514.0     &                       &  0.310  &  17.76  &  $-$23.35  &  4.67  &  30.04  &	  3.04    &     0.595  &	26.15  &   $-$1.495  &  $<0.41$  &   $<0.41$  &	 R  &   1747  &	       &  rs930729n00  \\
171821.03$+$564551.3     &                       &  0.619  &  22.35  &  $-$22.16  &  0.11  &  29.03  &  $<0.448$  &   $<0.108$ &     $<26.03$  &  $<-1.151$  &  $<0.41$  &   $<6.13$  &  H  &  16225  &   6.3  &  rh801039n00  \\
171845.14$+$562419.7     &                       &  1.422  &  19.50  &  $-$25.25  &  0.91  &  30.67  &  $<0.228$  &   $<0.083$ &     $<26.63$  &  $<-1.552$  &  $<0.42$  &   $<2.62$  &  H  &  16225  &  15.8  &  rh801039n00  \\
171901.69$+$564657.8     &                       &  3.920  &  20.19  &  $-$26.37  &  0.50  &  31.17  &  $<0.324$  &   $<0.238$ &     $<27.85$  &  $<-1.276$  &  $<0.44$  &   $<9.90$  &  H  &  16225  &   9.8  &  rh801039n00  \\
171901.70$+$580028.7     &                       &  0.774  &  19.23  &  $-$24.00  &  1.18  &  30.27  &  $<0.364$  &   $<0.096$ &     $<26.18$  &  $<-1.569$  &  $<0.44$  &   $<2.02$  &  P  &   5336  &  13.0  &  rp200724n00  \\
171902.29$+$593715.8     &                       &  0.179  &  18.23  &  $-$21.88  &  3.03  &  29.36  &	  4.88    &     0.858  &	25.82  &   $-$1.362  &  $<1.50$  &   $<1.71$  &	 R  &   2035  &	       &  rs930624n00  \\
171954.27$+$563103.2     &                       &  3.211  &  19.36  &  $-$26.72  &  1.00  &  31.33  &  $<0.260$  &   $<0.163$ &     $<27.55$  &  $<-1.454$  &  $<0.45$  &   $<4.65$  &  H  &  16225  &  16.2  &  rh801039n00  \\
172026.69$+$554024.1     &                       &  0.359  &  17.39  &  $-$24.05  &  6.57  &  30.33  &	  3.39    &     0.687  &	26.34  &   $-$1.528  &  $<0.44$  &   $<0.32$  &	 R  &   1287  &	       &  rs930729n00  \\
172032.28$+$551330.2     &                       &  0.273  &  18.50  &  $-$22.41  &  2.36  &  29.63  &	  2.94    &     0.558  &	26.01  &   $-$1.392  &  $<0.40$  &   $<0.71$  &	 R  &   1312  &	       &  rs930729n00  \\
172059.46$+$612811.8     &                       &  0.236  &  19.25  &  $-$21.68  &  1.18  &  29.20  &	  2.05    &     0.377  &	25.71  &   $-$1.342  &  $<1.50$  &   $<3.78$  &	 R  &   2016  &	       &  rs930624n00  \\
172100.76$+$632730.6     &                       &  0.464  &  20.00  &  $-$22.36  &  0.65  &  29.55  &	  0.937   &     0.205  &	26.05  &   $-$1.343  &  $<1.50$  &   $<9.46$  &	 R  &   2417  &	       &  rs930624n00  \\
172122.82$+$575029.4     &                       &  1.817  &  18.23  &  $-$26.79  &  2.79  &  31.35  &  $<0.240$  &   $<0.100$ &     $<26.91$  &  $<-1.706$  &  $<0.41$  &   $<1.09$  &  P  &   5336  &  11.4  &  rp200724n00  \\
172200.14$+$644357.3     &                       &  1.767  &  17.91  &  $-$27.05  &  3.56  &  31.44  &	  4.68    &     0.193  &	27.17  &   $-$1.637  &  $<1.50$  &   $<2.94$  &	 R  &   3697  &	       &  rs930521n00  \\
172358.01$+$601140.0     &                       &  2.239  &  19.15  &  $-$26.86  &  1.91  &  31.35  &	  0.806   &     0.390  &	27.66  &   $-$1.416  &  $<1.50$  &   $<6.06$  &	 R  &   2182  &	       &  rs930624n00  \\
172430.50$+$543952.4     &                       &  1.076  &  19.55  &  $-$24.99  &  1.14  &  30.54  &	  1.26    &     0.391  &	27.07  &   $-$1.330  &  $<0.46$  &   $<1.86$  &	 R  &   1499  &	       &  rs930729n00  \\
172446.40$+$603619.6     &                       &  0.372  &  18.97  &  $-$22.65  &  1.53  &  29.73  &	  1.01    &     0.206  &	25.85  &   $-$1.486  &  $<1.50$  &   $<4.33$  &	 R  &   2335  &	       &  rs930624n00  \\
172543.18$+$580604.8     &                       &  0.292  &  18.78  &  $-$22.42  &  1.82  &  29.58  &	  3.68    &     0.710  &	26.17  &   $-$1.309  &  $<0.44$  &   $<0.91$  &	 R  &   1638  &	       &  rs930729n00  \\
172823.60$+$630933.7     &                       &  0.439  &  19.18  &  $-$22.93  &  1.32  &  29.81  &	  0.670   &     0.144  &	25.85  &   $-$1.521  &  $<1.50$  &   $<4.91$  &	 R  &   2404  &	       &  rs930625n00  \\
172842.94$+$611057.1     &                       &  0.590  &  18.73  &  $-$23.98  &  1.92  &  30.24  &	  1.75    &     0.416  &	26.57  &   $-$1.407  &  $<1.50$  &   $<3.72$  &	 R  &   2538  &	       &  rs930625n00  \\
173203.10$+$611751.9     &                       &  0.358  &  19.31  &  $-$22.41  &  1.12  &  29.56  &	  1.07    &     0.216  &	25.84  &   $-$1.426  &  $<1.50$  &   $<4.98$  &	 R  &   2683  &	       &  rs930625n00  \\
173229.42$+$564811.3     &                       &  0.303  &  18.77  &  $-$22.50  &  1.84  &  29.62  &	  2.45    &     0.476  &	26.03  &   $-$1.377  &  $<0.44$  &   $<0.92$  &	 R  &   1883  &	       &  rs930729n00  \\
173348.81$+$585651.0     &                       &  0.491  &  18.12  &  $-$23.96  &  3.29  &  30.31  &	  2.13    &     0.475  &	26.47  &   $-$1.474  &  $<1.50$  &   $<2.48$  &	 R  &   2135  &	       &  rs930729n00  \\
173721.14$+$550321.6     &                       &  0.333  &  19.26  &  $-$22.41  &  1.17  &  29.51  &	  1.65    &     0.372  &	26.01  &   $-$1.343  &  $<0.44$  &   $<1.21$  &	 R  &   1643  &	       &  rs930729n00  \\
174146.49$+$540146.1     &                       &  0.416  &  19.64  &  $-$22.41  &  0.83  &  29.56  &	  1.85    &     0.391  &	26.23  &   $-$1.276  &  $<0.46$  &   $<2.14$  &	 R  &   1686  &	       &  rs930729n00  \\
233642.61$+$001653.3     &                       &  2.606  &  19.14  &  $-$26.64  &  1.20  &  31.26  &  $<0.350$  &   $<0.140$ &     $<27.33$  &  $<-1.510$  &  $<0.45$  &   $<3.14$  &  H  &  15215  &  14.0  &  rh800856n00  \\
233718.07$+$002550.6     &                       &  2.046  &  19.32  &  $-$26.09  &  1.20  &  31.08  &  $<0.405$  &   $<0.184$ &     $<27.27$  &  $<-1.464$  &  $<0.45$  &   $<3.03$  &  H  &  15215  &  10.9  &  rh800856n00  \\
233722.01$+$002238.8     &                       &  1.380  &  19.68  &  $-$25.13  &  0.97  &  30.67  &  $<0.478$  &   $<0.170$ &     $<26.92$  &  $<-1.442$  &  $<0.42$  &   $<2.71$  &  H  &  15215  &   7.6  &  rh800856n00  \\
233736.43$+$000950.7     &                       &  1.136  &  19.18  &  $-$24.89  &  1.16  &  30.59  &  $<0.329$  &   $<0.105$ &     $<26.55$  &  $<-1.552$  &  $<0.41$  &   $<2.06$  &  H  &  15215  &   6.4  &  rh800856n00  \\
233739.12$+$002656.2     &                       &  1.700  &  18.86  &  $-$26.11  &  1.62  &  31.07  &  $<0.441$  &   $<0.178$ &     $<27.11$  &  $<-1.520$  &  $<0.46$  &   $<1.93$  &  H  &  15215  &  10.7  &  rh800856n00  \\
234819.58$+$005721.4     &                       &  2.163  &  19.10  &  $-$26.19  &  1.10  &  31.08  &	  0.143   &     0.068  &	26.87  &   $-$1.616  &  $<0.42$  &   $<2.89$  &	 H  &  46315  &   0.0  &  rh701577a01  \\
234840.05$+$010753.5     &                       &  0.719  &  18.61  &  $-$24.27  &  2.13  &  30.46  &	  1.35    &     0.347  &	26.67  &   $-$1.454  &  $<0.45$  &   $<1.36$  &	 H  &  46315  &  11.6  &  rh701577a01  \\
235221.19$-$005125.9     &                       &  0.438  &  19.68  &  $-$22.32  &  0.98  &  29.59  &	  0.747   &     0.160  &	25.89  &   $-$1.419  &  $<0.45$  &   $<2.59$  &	 P  &   6867  &  18.5  &  rp701406n00  \\
\tableline
\enddata
\tablecomments{Luminosities are computed using $H_{0}$=70 km s$^{-1}$ Mpc$^{-1}$, 
$\Omega_{\rm M}$=0.3, and $\Omega_{\Lambda}$=0.7. 
The upper limits are at the 3$\sigma$ level.}
\tablenotetext{a}{Alternative name as reported in the SDSS EDR catalog.}
\tablenotetext{b}{Rest-frame flux density in units of 10$^{-27}$~erg cm$^{-2}$ s$^{-1}$ Hz$^{-1}$ (at 2500~\AA) and 
10$^{-30}$~erg cm$^{-2}$ s$^{-1}$ Hz$^{-1}$ (at 2~keV).}
\tablenotetext{c}{Rest-frame luminosity density (erg s$^{-1}$ Hz$^{-1}$).}
\tablenotetext{d}{Galactic absorption-corrected flux in the observed 0.5--2~keV band in units of 10$^{-13}$~erg cm$^{-2}$ s$^{-1}$.}
\tablenotetext{e}{Observed-frame 1.4~GHz flux density either from FIRST or NVSS (mJy).}
\tablenotetext{f}{Radio-loudness parameter, defined as 
$R$ = $f_{\rm 5~GHz}/f_{\rm 4400~\mbox{\tiny\AA}}$ (rest frame; e.g., Kellermann et al. 1989). 
The rest-frame 5~GHz flux density is computed from the observed-frame 1.4~GHz flux density 
assuming a radio power-law slope of $\alpha=-0.8$, with $f_{\nu}\propto~\nu^{\alpha}$.}
\tablenotetext{g}{X-ray instrument used in the analysis: 
R=\rosat\ All-Sky Survey; 
P=Pointed \rosat\ PSPC; 
H=Pointed \rosat\ HRI; 
C=\chandra\ ACIS-S;
X=\xmm\ EPIC.} 
\tablenotetext{h}{Exposure time (s).}
\tablenotetext{i}{Off-axis angle (arcmin).}
\tablenotetext{\dagger}{BALQSO.}
\label{tab2}
\end{deluxetable}
%%%%%%%%%%%%%%%
% aox TABLE End
%%%%%%%%%%%%%%%%%%%%%%%%%%%%%%%%%%%%%%%%%%

\end{document}